\documentclass[3p]{elsarticle}

\usepackage{amsmath,amsfonts,bm}

\def\eqref#1{equation~(\ref{#1})}
\def\Eqref#1{Equation~(\ref{#1})}
\def\plaineqref#1{(\ref{#1})}

\def\1{\bm{1}}

\DeclareMathAlphabet{\mathsfit}{\encodingdefault}{\sfdefault}{m}{sl}
\SetMathAlphabet{\mathsfit}{bold}{\encodingdefault}{\sfdefault}{bx}{n}

\DeclareMathOperator*{\argmin}{arg\,min}

\usepackage{hyperref}
\usepackage{breqn}
\usepackage{url}
\usepackage{graphicx}
\usepackage{amsmath}
\usepackage{enumitem}
\usepackage{caption}
\usepackage{subcaption}
\usepackage{caption}
\usepackage{booktabs}
\usepackage[export]{adjustbox}
\usepackage{multirow}
\usepackage{balance}
\usepackage{mdframed}
\usepackage{marginnote}
\usepackage{algpseudocode}
\usepackage{algorithm}

\usepackage{tcolorbox}

\usepackage{circledsteps}
            
\newcommand{\diag}{\text{diag}}
\newcommand{\tr}{\text{tr}}

\makeatletter
\def\ps@pprintTitle{%
 \let\@oddhead\@empty
 \let\@evenhead\@empty
 \def\@oddfoot{}%
 \let\@evenfoot\@oddfoot}
\makeatother

\hypersetup{
    colorlinks=true,
    linkcolor=blue,
    filecolor=magenta,      
    urlcolor=cyan
}

\biboptions{authoryear}

\begin{document}

\begin{frontmatter}
\title{Learning and controlling the source-filter representation of speech \\ with a variational autoencoder\tnoteref{t1}}

\author[1]{Samir Sadok\corref{cor1}}
\ead{samir.sadok@centralesupelec.fr}
\author[1]{Simon Leglaive\corref{cor1}}
\ead{simon.leglaive@centralesupelec.fr}
\author[2]{Laurent Girin}
\author[3]{Xavier Alameda-Pineda}
\author[1]{Renaud S\'eguier}
\address[1]{CentraleSupélec, IETR UMR CNRS 6164, France}
\address[2]{Univ.~Grenoble Alpes, CNRS, Grenoble-INP, GIPSA-lab, France}
\address[3]{Inria, Univ.~Grenoble Alpes, CNRS, LJK, France}
\cortext[cor1]{Corresponding authors}
\tnotetext[t1]{This research was supported by ANR-3IA MIAI (ANR-19-P3IA-0003), ANR-JCJC ML3RI (ANR-19-CE33-0008-01), H2020 SPRING (funded by EC under GA \#871245). }

\begin{abstract}
\label{abstract}
Understanding and controlling latent representations in deep generative models is a challenging yet important problem for analyzing, transforming and generating various types of data. In speech processing, inspiring from the anatomical mechanisms of phonation, the source-filter model considers that speech signals are produced from a few independent and physically meaningful continuous latent factors, among which the fundamental frequency $f_0$ and the formants are of primary importance. In this work, we start from a variational autoencoder (VAE) trained in an unsupervised manner on a large dataset of unlabeled natural speech signals, and we show that the source-filter model of speech production naturally arises as orthogonal subspaces of the VAE latent space. Using only a few seconds of labeled speech signals generated with an artificial speech synthesizer, we propose a method to identify the latent subspaces encoding $f_0$ and the first three formant frequencies, we show that these subspaces are orthogonal, and based on this orthogonality, we develop a method to accurately and independently control the source-filter speech factors within the latent subspaces. Without requiring additional information such as text or human-labeled data, this results in a deep generative model of speech spectrograms that is conditioned on $f_0$ and the formant frequencies, and which is applied to the transformation speech signals. Finally, we also propose a robust $f_0$ estimation method that exploits the projection of a speech signal onto the learned latent subspace associated with $f_0$.
\end{abstract}


\end{frontmatter}

\section{Introduction and related work}

\subsection{Introduction}

High-dimensional data such as natural images or speech signals exhibit some form of regularity which prevents their dimensions from varying independently from each other. This suggests that there exists a latent representation of smaller dimension from which the high-dimensional observed data are generated. Discovering such latent representation of complex data is the goal of representation learning, and deep latent-variable generative models have emerged as promising unsupervised approaches \citep{goodfellow2014generative, kingma2014auto, rezende2014stochastic, chen2016infogan, higgins2016beta, kim2018disentangling, chen2018isolating}. The variational autoencoder (VAE) \citep{kingma2014auto, rezende2014stochastic}, which is equipped with both a generative and inference model, can be used not only for data generation but also for analysis and transformation. As an explicit model of a probability density function (pdf), the VAE can also be used as a learned prior for solving inverse problems such as compressed sensing \citep{bora2017compressed}, speech enhancement \citep{bando2018statistical,leglaive_MLSP18}, or source separation \citep{kameoka2019supervised,jayaram2020source}. Making sense of the latent representation learned by a VAE and controlling the underlying continuous factors of variation in the data are important challenges to build more expressive and interpretable generative models and probabilistic priors.

A series of previous works on representation learning with deep generative models, in particular VAEs, have focused on images \citep{higgins2016beta,kim2018disentangling,chen2018isolating, locatello2019challenging, locatello2020sober}. Yet, it is not always easy to define the ground-truth latent factors of variation involved in the generation of natural images. For speech data, the latent factors of variation can be directly related to the anatomical mechanisms of speech production. Indeed, the source-filter model proposed by \cite{fant1970acoustic} considers that the production of speech signals results from the interaction of a source signal with a linear filter. In voiced speech, the source originates from the vibration of the vocal folds, which produces a quasi-periodic glottal sound wave whose fundamental frequency $f_0$, loosely referred to as the \emph{pitch}, is a key element of speech prosody. In unvoiced speech, a noise source is produced by a turbulent airflow or an acoustic impulse. The source signal is modified by the vocal tract, which is assumed to act as a linear filter. The cavities of the vocal tract give rise to resonances, which are called the \emph{formants} and are characterized by their frequency, amplitude, and bandwidth. By moving the speech articulators such as the tongue, lips, and jaw, humans modify the shape of their vocal tract, which results in a change of the acoustic filter, the associated resonances, and the resulting speech sounds. For voiced phonemes, humans are able to control the formants independently of the pitch, i.e., to change the filter independently of the source \citep{fant1970acoustic}, and of each other \citep{macdonald2011probing}. The source-filter model thus considers that a speech signal is mainly characterized by a few continuous latent factors of variation corresponding to the source, among which $f_0$ plays a central role, and to the filter, mostly characterized by the formants. The independence of the source and filter characteristics makes the speech signals an interesting material for disentangled representation learning methods, especially with deep generative latent-variable models such as the VAE. 

\begin{figure*}
    \centering
    \includegraphics[width=\linewidth]{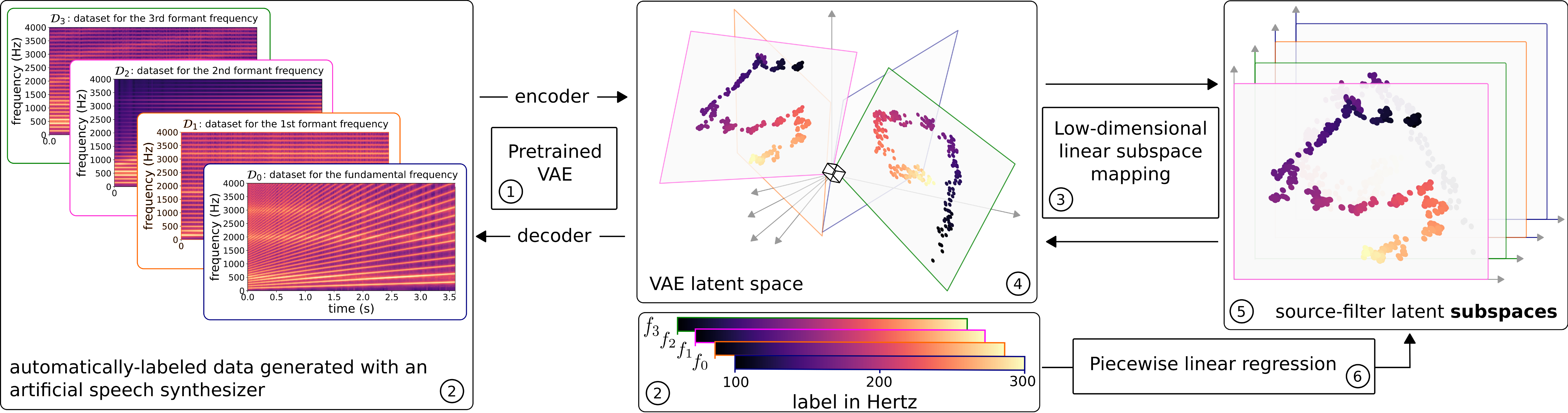}
    \caption{Overview of the proposed method. First, \Circled{1} a VAE is trained in an unsupervised manner by maximizing a lower bound of the data log-marginal likelihood (see Section~\ref{subsec:VAE}) on a large dataset of unlabeled natural speech signals (not shown on this figure for clarity). Given the pretrained VAE and given \Circled{2} a few seconds of automatically-labeled speech generated with an artificial speech synthesizer, we then propose \Circled{3} a linear subspace identification method to put in evidence that \Circled{4} the VAE latent space is structured into \Circled{5} orthogonal subspaces that encode $f_0$ and the formant frequencies, thus complying with the source-filter model of speech production. The subspaces are identified by minimizing the L2 norm of the reconstruction error obtained after passing the artificially-generated speech trajectories through the VAE encoder and projecting on the subspaces (see Section~\ref{subsec:learning_latent_subspaces}).
    Finally, we propose \Circled{6} a piecewise linear regression model to learn how to move into the source-filter latent subspaces, so as to perform speech manipulations in a disentangled manner. This model is also learned using the automatically-labeled artificial speech trajectories, by minimizing the L2 norm of the difference between the output of the regression model and the data coordinates in the previously-learned latent subspaces (see Section~\ref{subsec:control_factor}). No supervision is used to constrain the structure of the VAE latent space during its training. Supervision is only used after the training of the VAE, to identify the disentangled latent subspaces encoding the $f_0$ and formant frequencies, and to learn how to move into these subspaces to perform speech manipulations.}
    \label{fig:sfvae}
\end{figure*}

In this work, we analyze and control the latent space of a VAE from the perspective of the source-filter model of speech production, which can be beneficial for various applications in speech analysis, transformation, and synthesis. An overview of the proposed approach is shown in Figure~\ref{fig:sfvae}. We first train a VAE on a dataset of about 25~hours of unlabeled speech signals. Then, using only a few seconds of automatically labeled speech signals generated with an artificial speech synthesizer, we propose a method to identify and independently control the source-filter continuous latent factors of speech variation within the latent space of the VAE. Our contributions are the following: (i) We identify the source-filter model in the VAE latent space by showing experimentally that $f_0$ and the frequency of the first three formants, $f_1$, $f_2$, and $f_3$, are encoded in different subspaces. We put in evidence the orthogonality of the learned subspaces, which not only shows that the representation learned by the VAE complies with the source-filter model of speech production, but also suggests that we can perform speech transformations in a disentangled manner (i.e., modifying one of the factors would not affect the others) by moving into the learned subspaces. (ii) For each factor $f_i$, $i \in \{0, 1, 2, 3\}$, we propose to learn a piecewise linear regression model from the factor value in the synthetic speech dataset to the coordinates in the corresponding latent subspace. This method allows us to precisely and independently control the source-filter factors of speech variation within the learned subspaces, as confirmed experimentally on both artificial and natural signals. Without requiring additional information such as text or human-labeled data, the proposed approach leads to a deep generative model of speech spectrograms that is conditioned on $f_0$ and the formant frequencies. (iii) Finally, to illustrate the interest of the learned representation for downstream tasks, we propose an $f_0$ estimation method that exploits the projection of a speech signal onto the learned latent subspace associated with $f_0$. Experiments show that this approach competes with state-of-the-art methods in terms of precision and robustness to noise.

To the best of our knowledge, this is the first study showing the link between the classical source-filter model of speech production and the representation learned in the latent space of a VAE. Exploiting this link, we propose a principled method applied to the generation and transformation of speech signals controlled with interpretable trajectories of $f_0$ and the formant frequencies. Regarding this latter application, our objective is not to compete with traditional signal processing methods (these are discussed in the next subsection), which to the best of our knowledge remain the state-of-the-art. The interest of the present paper is rather to advance the understanding of deep generative modeling of speech signals while comparing fairly with traditional signal-model-based systems specifically designed for a given task. Moreover, advancing on the interpretability and control of the VAE latent space is expected to be beneficial for downstream tasks, for instance to develop pitch-informed extensions of VAE-based speech enhancement methods such as those of \citet{bando2018statistical, leglaive_MLSP18, leglaive2020recurrent, bie2021unsupervised}.

\subsection{Related work}
\label{sec:relatedwork}

Time-scale, pitch-scale, and timbre modification of speech signals is a highly covered research problem originally addressed with signal processing methods. Three main groups of approaches exist \citep{laroche2002time}: time-domain methods such as the pitch-synchronous overlap and add (PSOLA) algorithm \citep{moulines1990pitch}, methods that work in the short-time Fourier transform (STFT) domain such as the phase vocoder \citep{flanagan1966phase, laroche1999improved}, and parametric approaches based for instance on linear predictive coding (LPC) \citep{makhoul1975linear, markel1976linear}, sinusoidal modeling \citep{mcaulay1986speech, george1997speech}, or sinusoidal plus noise modeling \citep{serra1990spectral, laroche1993hns}. Other signal-processing-based approaches to real-time speech manipulations include the STRAIGHT \citep{kawahara2006straight, banno2007implementation} and WORLD \citep{morise2016world} vocoders, which exploit a decomposition of the speech signal into $f_0$, spectral envelope, and aperiodicity.

Deep learning has recently emerged as a powerful approach to speech signal manipulation. A few methods have investigated combining traditional signal processing models with deep learning \citep{valin2019lpcnet, wang2019neural, juvela2019glotnet, Lee2019AdversariallyTE, choi2021neural}.
LPCNet is a successful neural vocoder inspired by the source-filter model \citep{valin2019lpcnet}.
It was recently extended to pitch shifting and time stretching of speech signals by \citet{morrison2021neural}. Yet, the authors showed that time-domain PSOLA (TD-PSOLA) \citep{moulines1990pitch} remains a very strong baseline that is difficult to outperform with deep learning methods. 

Regarding the use of deep generative models (in particular VAEs) for speech modeling and transformation, the studies of \citet{blaauw2016modeling, hsu2016voice,HsuUnsupervised2017,hsu2017learning,akuzawa2018expressive} are pioneering. Of particular interest to the present paper is the work of \cite{hsu2017learning}. The authors proposed to use VAEs for modifying the speaker identity and the phonemic content of speech signals by translations in the latent space of a VAE. Yet, this method requires to know predefined values of the latent representations associated with both the source and target speech attributes to be modified. The performance of the method thus depends on the quality of the estimation of the source attribute (e.g., $f_0$), which has to be obtained from the input speech signal to be transformed. This differs from the proposed method which relies on projection onto the latent subspace associated with a given attribute, and only requires the target value for this attribute. Moreover, \citet{hsu2017learning} did not address the control of continuous factors of speech variation in the VAE latent space, contrary to the present work.

For deep latent representation learning methods, the challenge is to relate the learned representation to interpretable speech attributes. In \citet{qian2020unsupervised} and \citet{webber2020hider}, this interpretability is enforced by the design of the model. \citet{qian2020unsupervised} proposed to use three independent encoder networks to decompose a speech signal into $f_0$, timbre and rhythm latent representations. \citet{webber2020hider} focused on controlling source-filter parameters in speech signals, where the ability to control a given parameter (e.g., $f_0$) is enforced explicitly using labeled data and adversarial learning. In this approach, each parameter to be controlled requires dedicated training of the model. Moreover, these methods are speaker-dependent, as speech generation in \citet{qian2020unsupervised} is conditioned on the speaker identity and \citet{webber2020hider} used a single-speaker training dataset. This contrasts with the proposed method which is speaker-independent, and in which the source-filter representation is shown to emerge as orthogonal subspaces of the latent space of a single unsupervised VAE model.

In the machine learning and computer vision communities, variants of the VAE have recently led to considerable progress in disentangled representation learning \citep{kim2018disentangling, higgins2016beta, chen2018isolating}. From experimental analyses on image data, these methods suggest that a vanilla VAE cannot learn a disentangled representation. Moreover, \cite{locatello2019challenging, locatello2020sober} recently showed both theoretically and from a large-scale experimental study that the unsupervised learning of disentangled representations is impossible without inductive biases (i.e., implicit or explicit assumptions by which a machine learning algorithm is able to generalize) on both the models and the data.
Weakly-supervised \citep{hosoya2018group, shu2019weakly, locatello2020weakly} and semi-supervised  \citep{locatello2019disentangling, sorrenson2020disentanglement} methods have thus been proposed to learn disentangled representations. 
For example, the semi-supervised approach of \cite{locatello2019disentangling} exploits a small amount of labelled data to enforce the disentanglement of the representation at training time.
This differs from the proposed approach where, after training a VAE on unlabeled natural speech signals, a few examples of artificially-generated labeled speech data are used to identify the disentangled structure of the VAE latent representation in terms of source-filter factors of speech variation. This allows us to experimentally show that learning a disentangled source-filter representation of speech using a simple VAE is possible, complying with the definition of disentanglement proposed in \citep{higgins2018towards}.

Several methods have been recently proposed to control continuous factors of variation in deep generative models \citep{jahanian2019steerability, Plumerault2020Controlling, goetschalckx2019ganalyze,harkonen2020ganspace}, focusing essentially on generative adversarial networks \citep{goodfellow2014generative}. They consist in identifying and then moving onto semantically meaningful directions in the latent space of the model. The present work is inspired by \citep{Plumerault2020Controlling}, which assumes that a factor of variation can be predicted from the projection of the latent vector along a specific axis, learned from artificially generated trajectories. The proposed method is however more generic, thanks to the learning of latent subspaces associated with the latent factors and to the introduction of a general formalism based on the use of ``biased’’ aggregated posteriors. Moreover, the previous works on controlling deep generative models only allow for moving ``blindly’’ onto semantically meaningful directions in the latent space. In the present study, we are able to generate data conditioned on a specific target value for a given factor of variation (e.g., a given formant frequency value). Finally, previous works focused on image data. To the best of our knowledge, the present paper proposes the first approach to identify and control source-filter factors of speech variation in a VAE.

The rest of this paper is organized as follows: Section~\ref{sec:method} presents the proposed method for analyzing and controlling source-filter factors of speech variation in a VAE. The method is evaluated experimentally and compared with traditional signal processing algorithms and with the approach of \cite{hsu2017learning} in Section~\ref{sec:results}. We finally conclude in Section~\ref{sec:conclusion}.

\section{Analyzing and controlling source-filter factors of speech variation in a VAE}
\label{sec:method}
In this section, we first present the VAE model that we build upon, and that is trained on a large dataset of natural unlabeled speech signals. Then, we present (i) the method that we propose to identify latent subspaces encoding source-filter factors of speech variation from a few seconds of artificially-generated labeled speech signals, (ii) a simple strategy to measure the disentanglement of the learned representation, (iii) a method to control the continuous factors of variation in the learned subspaces and generate corresponding speech signals, and (iv) a simple method to estimate the $f_0$ contour of a speech signal by exploiting its projection onto the corresponding latent subspace.

\subsection{Variational autoencoder}
\label{subsec:VAE}

Generative modeling consists in learning a probabilistic model of an observable random variable $\mathbf{x} \in \mathcal{X} \subset \mathbb{R}^D$. Let $\mathcal{D} = \{\mathbf{x}_1, ..., \mathbf{x}_N \in \mathcal{X}\}$ be a dataset of $N = \# \mathcal{D}$ independent and identically distributed (i.i.d.) observations of $\mathbf{x}$, where $\#\mathcal{D}$ denotes the cardinal of $\mathcal{D}$. The empirical distribution of $\mathbf{x}$ is defined by $\hat{p}(\mathbf{x}) = \frac{1}{N}\sum_{\mathbf{x}_n \in \mathcal{D}} \delta(\mathbf{x} - \mathbf{x}_n)$,
where $\delta$ is the Dirac delta function, which is null everywhere except in 0 where it takes the value 1.

The variational autoencoder (VAE) \citep{kingma2014auto, rezende2014stochastic} attempts to approximate $\hat{p}(\mathbf{x})$ with a pdf $p_\theta(\mathbf{x})$ parametrized by $\theta$. High-dimensional data such as natural images or speech signals exhibit some form of regularity which prevents the $D$ dimensions of $\mathbf{x}$ from varying independently from each other. We can thus assume that there exists a latent variable $\mathbf{z} \in \mathbb{R}^L$, with $L \ll D$, from which the observed data were generated. Accordingly, the model distribution in the VAE is defined by marginalizing the joint distribution of the latent and observed data, $p_\theta(\mathbf{x}) = \int p_\theta(\mathbf{x} | \mathbf{z}) p(\mathbf{z}) d\mathbf{z}$.

In this work, the observed data vector $\mathbf{x} \in \mathcal{X} = \mathbb{R}_+^D$ denotes the power spectrum of a short frame of speech signal (i.e., a column of the short-time Fourier transform (STFT) power spectrogram). Its entries are thus non-negative and its dimension $D$ equals the number of frequency bins. We use the Itakura-Saito VAE (IS-VAE) \citep{bando2018statistical,leglaive_MLSP18, girin2019} defined by
\begin{align}
    p(\mathbf{z}) &= \mathcal{N}(\mathbf{z}; \mathbf{0}, \mathbf{I}), \\ 
    p_\theta(\mathbf{x} | \mathbf{z}) &= \prod\limits_{d=1}^{D} \text{Exp}\Big( [\mathbf{x}]_d ; [\mathbf{v}_\theta(\mathbf{z})]_d^{-1} \Big), \label{vae_gen_dist}
\end{align}
where $\mathcal{N}$ and $\text{Exp}$ denote the densities of the multivariate Gaussian and univariate exponential distributions, respectively, and $[\mathbf{v}]_d$ denotes the $d$-th entry of $\mathbf{v}$. The inverse scale parameters of $p_\theta(\mathbf{x} | \mathbf{z})$ are provided by a neural network called the decoder, parametrized by $\theta$ and taking $\mathbf{z}$ as input.

The marginal likelihood $p_\theta(\mathbf{x})$ and the posterior distribution $p_\theta(\mathbf{z} | \mathbf{x})$ are intractable due to the nonlinearities of the decoder, so it is necessary to introduce an inference model $q_\phi(\mathbf{z} | \mathbf{x}) \approx p_\theta(\mathbf{z} | \mathbf{x})$, which is defined by
\begin{align}
     q_\phi(\mathbf{z} | \mathbf{x}) &= \mathcal{N}\left(\mathbf{z}; \boldsymbol{\mu}_{\phi}(\mathbf{x}), \diag\{\mathbf{v}_{\phi}(\mathbf{x})\} \right), \label{vae_inf_dist}
\end{align}
where the mean and variance parameters are provided by a neural network called the encoder network, parametrized by $\phi$ and taking $\mathbf{x}$ as input. Then, the VAE training consists in maximizing a lower-bound of $\ln p_\theta(\mathbf{x})$, called the evidence lower-bound (ELBO) and defined by
\begin{equation}
    \mathcal{L}(\theta, \phi) = \mathbb{E}_{\hat{p}(\mathbf{x})}\Big[  \mathbb{E}_{q_\phi(\mathbf{z} | \mathbf{x})} \big[ p_\theta(\mathbf{x} | \mathbf{z}) \big] - D_{\text{KL}}\big(q_\phi(\mathbf{z} | \mathbf{x}) \parallel p(\mathbf{z}) \big)\Big],
\end{equation}
where $D_{\text{KL}}\big(q \parallel p \big) = \mathbb{E}_{q} \left[ \ln q - \ln p \right]$ is the Kullback-Leibler divergence. During training, the generative and inference model parameters $\theta$ and $\phi$ are jointly estimated by maximizing the ELBO, using (variants of) stochastic gradient descent with the so-called reparameterization trick \citep{kingma2014auto, rezende2014stochastic}. 

\subsection{Learning source-filter latent subspaces}
\label{subsec:learning_latent_subspaces}

In addition to the pre-trained IS-VAE speech spectrogram model introduced in the previous subsection, we also assume the availability of an artificial speech synthesizer allowing for an accurate and independent control of $f_0$ and the formant frequencies. We use Soundgen \citep{Anikin2019}, a parametric synthesizer based on the source-filter model of speech production. For a given speech sound, the voiced component of the source signal is generated by a sum of sine waves, the noise component by a filtered white noise, and both components are then summed and passed through a linear filter simulating the effect of the human vocal tract. Importantly, this synthesizer allows us to easily generate artificial speech data labeled with $f_0$ and formant frequency values.

Formally, let $f_i$ denote the speech factor of variation (in Hz) corresponding to the fundamental frequency, for $i=0$, and to the formant frequencies, for $i \in \{ 1, 2, 3 \}$.
Let $\mathcal{D}_i$ denote a dataset of artificially-generated speech vectors (more precisely short-term power spectra) synthesized by varying only $f_i$, all other factors $\{f_j, j \neq i\}$ being arbitrarily fixed. All examples in $\mathcal{D}_i$ are labeled with the index and the value of the factor of variation. It would be relatively difficult to build such a dataset from existing corpora of unlabeled natural speech. In contrast, it is a very easy task using an artificial speech synthesizer such as Soundgen \citep{Anikin2019}, which precisely takes $f_0$ and the formant parameters as input, and outputs waveforms from which we extract power spectra. 

Let $\hat{p}^{(i)}(\mathbf{x})$ denote the empirical distribution associated with $\mathcal{D}_i$, defined similarly as $\hat{p}(\mathbf{x})$. We also introduce the following marginal distribution over the latent vectors:
\begin{equation}
    \hat{q}_\phi^{(i)}(\mathbf{z}) = \int q_\phi(\mathbf{z} | \mathbf{x}) \hat{p}^{(i)}(\mathbf{x}) d\mathbf{x} = \frac{1}{\# \mathcal{D}_i} \sum_{\mathbf{x}_n \in \mathcal{D}_i} q_\phi(\mathbf{z} | \mathbf{x}_n).
    \label{agg_posterior}
\end{equation}
In the literature, this quantity is referred to as the aggregated posterior \citep{makhzani2015adversarial}. However, $q_\phi(\mathbf{z} | \mathbf{x})$ is usually aggregated over the empirical distribution $\hat{p}(\mathbf{x})$ such that the aggregated posterior is expected to match with the prior $p(\mathbf{z})$ \citep{chen2018isolating, dai2018diagnosing}. In contrast, in \Eqref{agg_posterior} we aggregate over the ``biased'' data distribution $\hat{p}^{(i)}(\mathbf{x})$, where we know only one latent factor varies. This defines the explicit inductive bias \citep{locatello2019challenging} that we exploit to learn the latent source-filter representation of speech in the VAE.

In the following of the paper, without loss of generality, we assume that, for each data vector in $\mathcal{D}_i$, the associated latent vector $\mathbf{z}$ has been centered by subtracting the mean vector
\begin{equation}
    \boldsymbol{\mu}_\phi(\mathcal{D}_i) = \mathbb{E}_{\hat{q}_\phi^{(i)}(\mathbf{z})}[\mathbf{z}] = \frac{1}{\# \mathcal{D}_i}\sum_{\mathbf{x}_n \in \mathcal{D}_i} \boldsymbol{\mu}_{\phi}(\mathbf{x}_n).
    \label{mu_Di}
\end{equation}
Because only one factor varies in $\mathcal{D}_i$, we expect latent vectors drawn from the ``biased'' aggregated posterior in \Eqref{agg_posterior} to live on a low-dimensional manifold embedded in the original latent space $\mathbb{R}^L$. We assume this manifold to be a subspace characterized by its semi-orthogonal basis matrix $\mathbf{U}_i \in \mathbb{R}^{L \times M_i}$, $1 \le M_i < L$. This matrix is computed by solving the following optimization problem:
\begin{align}
\min_{\mathbf{U} \in \mathbb{R}^{L \times M_i}}\,\,  \mathbb{E}_{\hat{q}_\phi^{(i)}(\mathbf{z})} \left[ \left\lVert \mathbf{z} - \mathbf{U} \mathbf{U}^\top \mathbf{z} \right\rVert_2^2 \right], \quad s.t.\,\, \mathbf{U}^\top \mathbf{U} = \mathbf{I}.
\label{pca_opt_prb_1}
\end{align}
The space spanned by the columns of $\mathbf{U}_i$ is a subspace of the original latent space $\mathbb{R}^L$ in which the latent vectors associated with the variation of the factor $f_i$ in $\mathcal{D}_i$ are expected to live.
In \ref{app:PCA}, we show that, similarly to the  principal component analysis (PCA) \citep{pearson1901liii}, the solution to the optimization problem~\plaineqref{pca_opt_prb_1} is given by the $M_i$ eigenvectors corresponding to the $M_i$ largest eigenvalues~of
\begin{align}
     \mathbf{S}_\phi(\mathcal{D}_i) =& \frac{1}{\# \mathcal{D}_i} \sum_{\mathbf{x}_n \in \mathcal{D}_i} \Big[ \boldsymbol{\mu}_{\phi}(\mathbf{x}_n)\boldsymbol{\mu}_{\phi}(\mathbf{x}_n)^\top + \diag\{\mathbf{v}_{\phi}(\mathbf{x}_n)\} \Big] - \boldsymbol{\mu}_\phi(\mathcal{D}_i)\boldsymbol{\mu}_\phi(\mathcal{D}_i)^\top.
    \label{S_Di}   
\end{align}
The dimension $M_i$ of the subspace can be chosen such as to retain a certain percentage of the data variance in the latent space. Note that the only source of supervision used here is the knowledge that only the factor $f_i$ varies in the dataset $\mathcal{D}_i$.

\subsection{Disentanglement analysis of the latent representation}
\label{subsec:orthogonality}
As defined by \cite{higgins2018towards}, a representation is disentangled if it is possible to learn orthogonal latent subspaces associated with each factor of variation, whether they are single- or multi-dimensional. The approach presented in the previous subsection exactly follows this definition and offers a natural and straightforward way to objectively measure if the unsupervised VAE managed to learn a disentangled representation of the factors of variation under consideration. First, by simply looking at the eigenvalues associated with the columns of $\mathbf{U}_i \in \mathbb{R}^{L \times M_i}$, we can measure the amount of variance that is retained by the projection $\mathbf{U}_i \mathbf{U}_i^\top$. If a small number of components $M_i$ represents most of the variance, it indicates that only a few intrinsic dimensions of the latent space are dedicated to the factor of variation $f_i$ and varying this factor can be done by affine transformations. Second, if for two different factors of variation $f_i$ and $f_j$, with $i\neq j$, the columns of $\mathbf{U}_i$ are orthogonal to those of $\mathbf{U}_j$, this indicates that the two factors are encoded in orthogonal subspaces and therefore disentangled. It should however be verified experimentally that applying transformations by moving onto the subspace associated with $f_i$ generalizes to values of $\{f_j, j \neq i\}$ different than the ones used in $\mathcal{D}_i$.  

\subsection{Controlling the source-filter factors of variation}
\label{subsec:control_factor}

So far, for each factor $f_i$, we have defined a methodology to learn a latent subspace $\mathbf{U}_i \in \mathbb{R}^{L \times M_i} $ that encodes its variations in the dataset $\mathcal{D}_i$, containing a few examples of speech data generated by an artificial synthesizer. Making now use of the values of the factor $f_i$ for the data in $\mathcal{D}_i$, we learn a regression model $\mathbf{g}_{\eta_i}: \mathbb{R}_+ \mapsto \mathbb{R}^{M_i}$ from $f_i$, whose value is denoted by $y \in \mathbb{R}_+$, to the data coordinates in the latent subspace defined by $\mathbf{U}_i$. The parameters $\eta_i$ are thus defined as the solution of the following optimization problem:
\begin{align}
    & \min_\eta\,\, \Bigg\{ \mathbb{E}_{\hat{q}_\phi^{(i)}(\mathbf{z} , y)} \left[ \left\lVert \mathbf{g}_{\eta}(y) - \mathbf{U}_i^\top \mathbf{z} \right\rVert_2^2 \right] \overset{c}{=} \frac{1}{\# \mathcal{D}_i} \sum_{(\mathbf{x}_n, y_n) \in \mathcal{D}_i} \left\lVert \mathbf{g}_{\eta}(y_n) - \mathbf{U}_i^\top \big(\boldsymbol{\mu}_{\phi}(\mathbf{x}_n) - \boldsymbol{\mu}_{\phi}(\mathcal{D}_i)\big) \right\rVert_2^2 \Bigg\}, 
    \label{reg_opt_prb_1}
\end{align}
where $\hat{q}_\phi^{(i)}(\mathbf{z} , y) = \int q_\phi(\mathbf{z} | \mathbf{x}) \hat{p}^{(i)}(\mathbf{x}, y) d\mathbf{x} $, $\hat{p}^{(i)}(\mathbf{x}, y)$ is the empirical distribution associated with $\mathcal{D}_i$, considering now both the speech data vector $\mathbf{x}$ and the value $y$ of $f_i$, and $\overset{c}{=}$ denotes equality up to an additive constant w.r.t.~$\eta$. This approach can be seen as a probabilistic extension of principal component regression \citep{hotelling1957relations, Kendall}. 
For simplicity and because it revealed efficient for this task, $\mathbf{g}_{\eta_i}$ is chosen as a piece-wise linear regression model learned independently for each output coordinate $m \in \{1,...,M_i\}$. This choice is supported by the fact that the semi-orthogonal matrix $\mathbf{U}_i$ decorrelates the data \citep{bengio2013representation}. Solving the optimization problem~\plaineqref{reg_opt_prb_1} then consists in solving a linear system of equations. In this work, we used the Python library of \citet{jekel2019pwlf}. 
The learning of the regression models is summarized in Algorithm~\ref{alg:regression} of \ref{app:algorithms}.
It is important to remind that even if the regression model is supervised, the labeled dataset $\mathcal{D}_i$ is very small with only a few seconds of speech signals (see experimental setup details in \ref{app:setup}), it is synthetic, and the values of $f_i$ are automatically obtained during the generation of the data with an artificial speech synthesizer, so no manual annotation effort is required.

We can now transform a speech spectrogram by analyzing it with the VAE encoder, then linearly moving in the learned subspaces using the above regression model, and finally resynthesizing it with the VAE decoder.
Given a source latent vector $\mathbf{z}$ and a target value $y$ for the factor $f_i$, we apply the following affine transformation:
\begin{equation}
    \tilde{\mathbf{z}} = \mathbf{z} - \mathbf{U}_i \mathbf{U}_i^\top \mathbf{z} + \mathbf{U}_i \mathbf{g}_{\eta_i}(y).
    \label{z_tilde}
\end{equation}
This transformation consists in (i) subtracting the projection of $\mathbf{z}$ onto the subspace associated with the factor of variation $f_i$; and (ii) adding the target component provided by the regression model $\mathbf{g}_{\eta_i}$ mapped from the learned subspace to the original latent space by the matrix $\mathbf{U}_i$. This operation allows us to move only in the latent subspace associated with the factor $f_i$. If this subspace is orthogonal to the latent subspaces associated with the other factors $\{f_j, j \neq i\}$, the latter should remain the same between $\mathbf{z}$ and $\tilde{\mathbf{z}}$, only $f_i$ should be modified. This process can be straightforwardly generalized to multiple factors, by subtracting and adding terms corresponding to each one of them. Contrary to \citet{hsu2017learning}, the operation in \Eqref{z_tilde} does not require the knowledge of the factor $f_i$ value associated with the source vector $\mathbf{z}$, it only requires the value $y$ of the factor $f_i$ corresponding to the target vector $\tilde{\mathbf{z}}$ (this value $y$ being used as input to the regression model.)

Finally, as the prior $p(\mathbf{z})$ and inference model $q_\phi(\mathbf{z} | \mathbf{x})$ are Gaussian (see Equations~\plaineqref{vae_gen_dist} and \plaineqref{vae_inf_dist}), the transformation in \Eqref{z_tilde} has the following probabilistic formulation (using $\mathbf{U}_i^\top\mathbf{U}_i = \mathbf{I}$):

\begin{align}
    p(\tilde{\mathbf{z}} ; f_i = y) &= \mathcal{N}\Big(\tilde{\mathbf{z}}; \mathbf{U}_i \mathbf{g}_{\eta_i}(y), \mathbf{M}_i \Big) \label{conditional_prior} \\
    q_\phi(\tilde{\mathbf{z}} | \mathbf{x} ; f_i = y) &= \mathcal{N}\Big(\tilde{\mathbf{z}}; \mathbf{U}_i \mathbf{g}_{\eta_i}(y) + \mathbf{M}_i\boldsymbol{\mu}_{\phi}(\mathbf{x}), \mathbf{M}_i \diag\{\mathbf{v}_{\phi}(\mathbf{x})\} \Big),
    \label{eq:q-z-tilde}
\end{align}
where $\mathbf{M}_i = \mathbf{I} - \mathbf{U}_i\mathbf{U}_i^\top$. The prior in \Eqref{conditional_prior} is now conditioned on the factor $f_i$ and can be used to generate speech data given input trajectories of $f_0$ and formant frequencies. As we assumed centered latent data, the mean vector $\boldsymbol{\mu}_\phi(\mathcal{D}_i)$ defined in \Eqref{mu_Di} must be added to $\tilde{\mathbf{z}}$ before mapping this vector through the generative model $p_\theta(\mathbf{x} | \mathbf{z})$. The proposed method to manipulate the $f_0$ and formant frequencies in the latent source-filter subspaces is summarized in Algorithm~\ref{alg:control} of \ref{app:algorithms}.

\subsection{Estimating the fundamental frequency using the learned latent representation}
\label{subsec:analyse_factor}

To illustrate the interest of the learned representation on an analysis task, we propose to estimate the $f_0$ contour of a speech signal using its projection onto the corresponding latent subspace characterized by the estimated matrix $\mathbf{U}_0 \in \mathbb{R}^{L \times M_0}$ (cf. Section~\ref{subsec:learning_latent_subspaces}). As we focus on the analysis of $f_0$, in this subsection we assume that the latent vectors are centered by subtracting the mean vector $\boldsymbol{\mu}_\phi(\mathcal{D}_0)$ defined in \Eqref{mu_Di}: $\mathbf{z} \leftarrow \mathbf{z} - \boldsymbol{\mu}_\phi(\mathcal{D}_0)$. Let $\mathbf{p} = \mathbf{U}_0^\top \mathbf{z} \in \mathbb{R}^{M_0}$ denote the projection of $\mathbf{z}$ onto the $f_0$ latent subspace.\footnote{The term projection used to refer to $\mathbf{p} \in \mathbb{R}^{M_0}$ is a misuse of language. Strictly speaking, the projection of $\mathbf{z} \in \mathbb{R}^{L}$ onto the subspace characterized by $\mathbf{U}_0 \in \mathbb{R}^{L \times M_0}$ is given by $\mathbf{U}_0\mathbf{U}_0^\top \mathbf{z} \in \mathbb{R}^{L}$.}
Because $\mathbf{p}$ results from a linear transformation of $\mathbf{z}$ and the approximate posterior distribution $q_\phi(\mathbf{z} | \mathbf{x})$ defined in \eqref{vae_inf_dist} is Gaussian, we have:
\begin{equation}
    q_\phi(\mathbf{p} | \mathbf{x}) = \mathcal{N}\left(\mathbf{p}; \mathbf{U}_0^\top\boldsymbol{\mu}_\phi(\mathbf{x}),
    \mathbf{U}_0^\top  \diag\{\mathbf{v}_\phi(\mathbf{x})\} \mathbf{U}_0\right).
    \label{approx_posterior_p}
\end{equation}
As will be confirmed experimentally in Section~\ref{sec:results}, the subspace generated by $\mathbf{U}_0$ encodes the fundamental frequency information and the formant frequencies are encoded in other orthogonal subspaces. Therefore, the projection of $\mathbf{z}$ onto $\mathbf{U}_0$ is expected to provide invariance to a change of the formant frequencies, which is an appealing feature for estimating the fundamental frequency.
The method we propose is simple but effective as will be shown experimentally. For an input speech power spectrum $\mathbf{x}^{\text{test}}$ assumed to be voiced, the estimated fundamental frequency is given by the value $y$ of $f_0$ associated with $\mathbf{x}^\star \in \mathcal{D}_0$ defined by 
\begin{equation}
\mathbf{x}^\star = \argmin_{\mathbf{x} \in \mathcal{D}_0}  D_{\text{KL}}\left(q_\phi(\mathbf{p} | \mathbf{x}^{\text{test}}) \parallel q_\phi(\mathbf{p} | \mathbf{x}) \right).
\label{KL_f0_est}
\end{equation}
Using the KL divergence allows us to base the estimation of $f_0$ on the full distribution of the projection, i.e. taking not only the mean of the projection in \Eqref{approx_posterior_p} into account, but also the covariance.
The proposed method requires computing the KL divergence between two multivariate Gaussians, which admits a closed-form expression.
We can thus simply compute the KL divergence in \Eqref{KL_f0_est} numerically for all the examples $\mathbf{x}$ in the synthetic labeled dataset $\mathcal{D}_0$ and return the value $y$ of $f_0$ associated with the minimum. 

The $f_0$ estimation is done independently for each frame of the power spectrogram of an input speech signal. The resulting ``raw'' estimated $f_0$ trajectory is then smoothed by applying a median filter with a window size of 5 frames. Above, we assumed $\mathbf{x}^{\text{test}}$ was a voiced speech spectrum. 
The voiced/unvoiced detection can be made automatically by setting a threshold on the minimum value of the above KL divergence, i.e. $D_{\text{KL}}\left(q_\phi(\mathbf{p} | \mathbf{x}^{\text{test}}) \parallel q_\phi(\mathbf{p} | \mathbf{x}^\star)\right)$.

\section{Experiments}
\label{sec:results}

This section presents qualitative and quantitative experimental results obtained with the proposed VAE-based method for controlling $f_0$ and the formant frequencies of speech signals. The VAE is trained on about 25 hours of multi-speaker speech data from the Wall Street Journal (WSJ0) dataset \citep{WSJ0}. The data space dimension is $513$ and the latent space dimension is set to $16$. This dimension was chosen based on previous work showing it is optimal for the modeling of speech power spectra in the context of speech enhancement \citep{leglaive_MLSP18, Leglaive_ICASSP2019a, sekiguchi2019semi}. 
For a given factor of variation, the corresponding latent subspace is learned (see Section~\ref{subsec:learning_latent_subspaces}) using short trajectories of speech power spectra  (corresponding to a few seconds of speech) generated with Soundgen \citep{Anikin2019}, all other factors being arbitrarily fixed. When solving the optimization problem~\plaineqref{pca_opt_prb_1}, the latent subspace dimension $M_i$ of each factor of variation is chosen such that 80\% of the data variance is retained. This leads $M_0=4$, $M_1=1$ and $M_2 = M_3 = 3$. The regression models used to control the speech factors of variation in the latent space (see Section~\ref{subsec:control_factor}) are learned on the same trajectories, but using the values of the Soundgen input control parameters (i.e., $f_0$ and formant frequencies values). More details on the experimental set-up can be found in \ref{app:setup}. Given a generated or transformed spectrogram, we use Waveglow \citep{prenger2019waveglow} to reconstruct the time-domain signal. 

\begin{figure*}[t]
     \centering
     \begin{subfigure}[b]{0.49\textwidth}
         \centering
         \includegraphics[width=\textwidth]{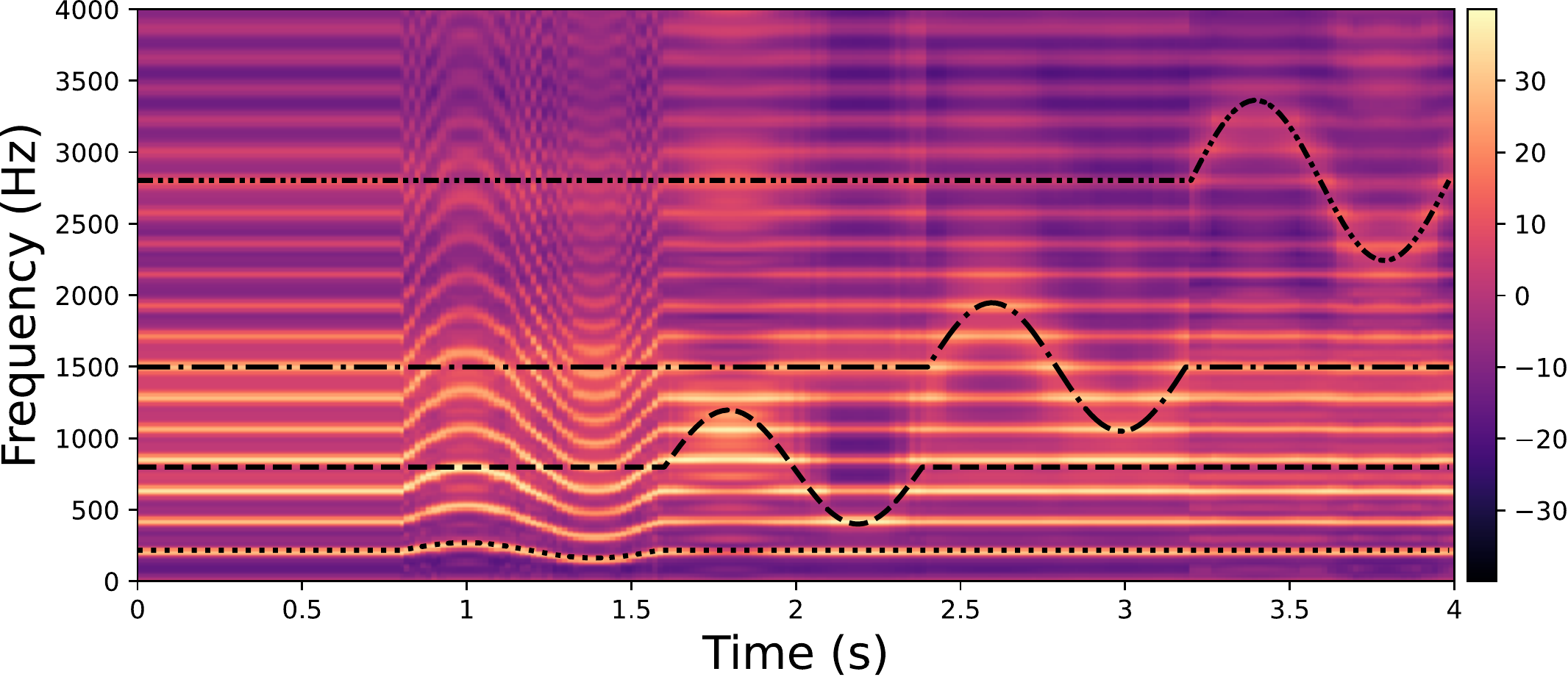}
         \label{fig:trans_spec}
     \end{subfigure}
     \begin{subfigure}[b]{0.49\textwidth}
         \centering
         \includegraphics[width=\textwidth]{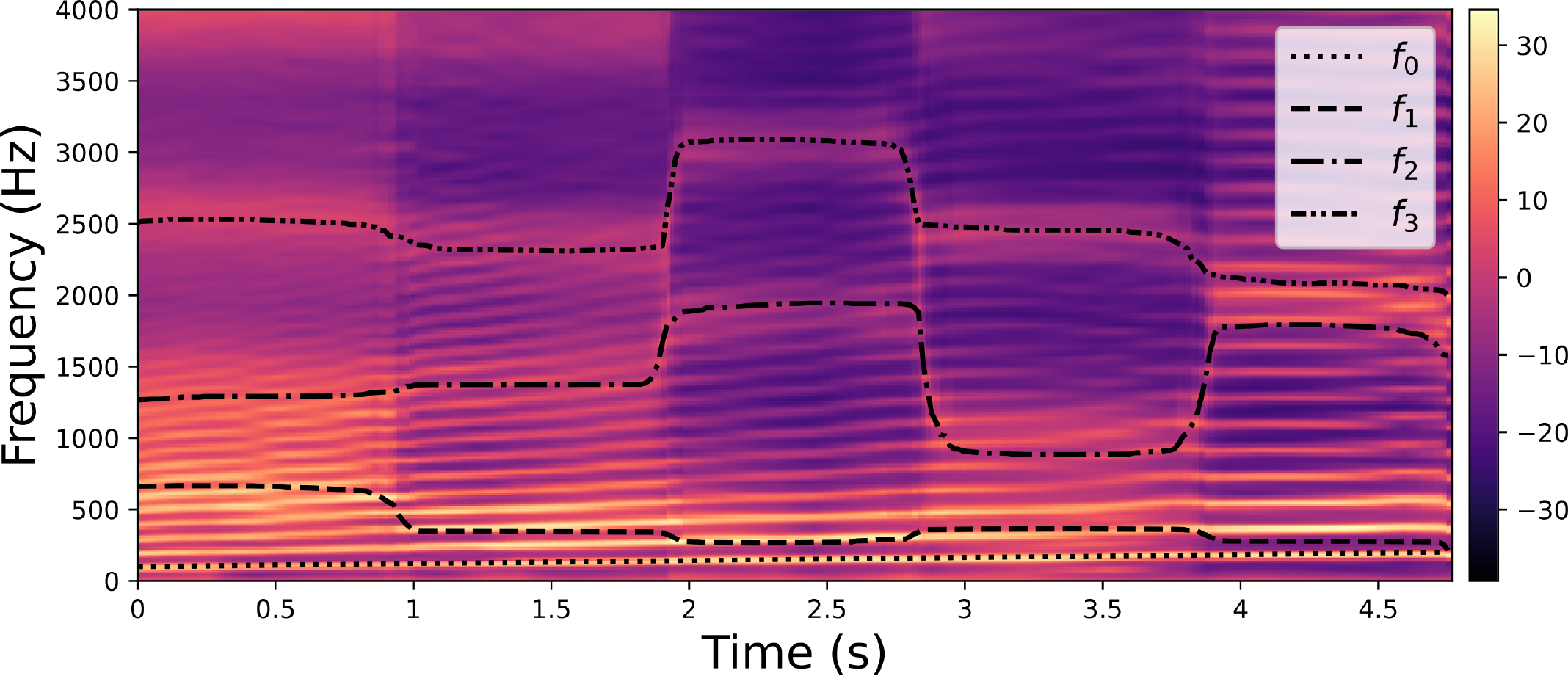}
         \label{fig:gen_spec}
     \end{subfigure}
     
     \begin{subfigure}[b]{\textwidth}
         \centering
         \includegraphics[width=\textwidth]{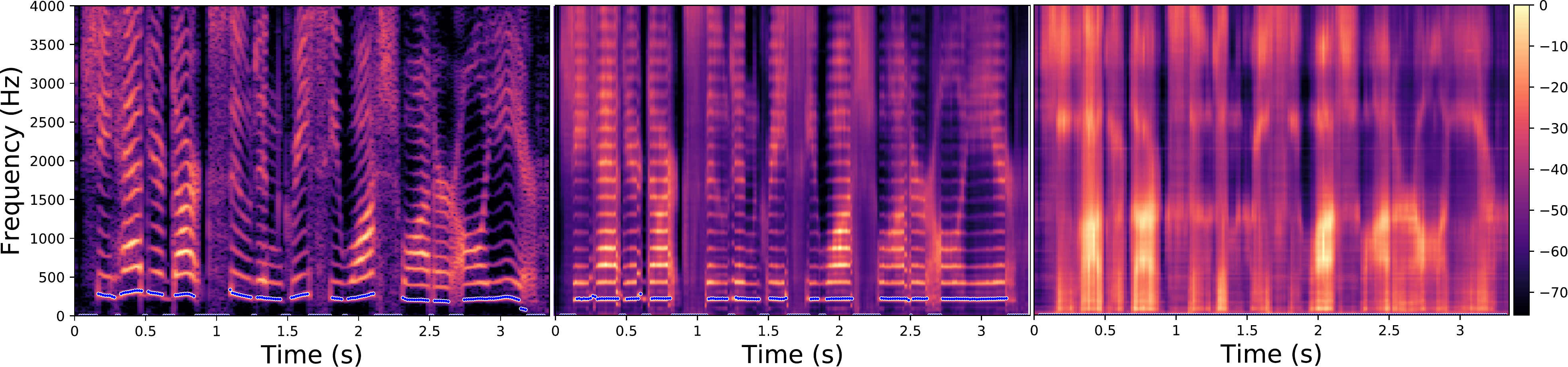}
         \label{fig:ori_mod_spec}
     \end{subfigure} 
     
     \caption{Examples of spectrograms modified and generated with the proposed method. The color bar indicates the power in dB. Top left: $f_0$ and formant transformations of a vowel /a/ uttered by a female speaker. Top right: Spectrogram generated from input trajectories of $f_0$ and formant frequencies. The target values of the factors $f_i$ are indicated by the black lines. Bottom left: Original spectrogram of a speech signal uttered by a female speaker; Bottom middle: Transformed spectrogram with $f_0$ (blue line) set constant over time; Bottom right: Transformed spectrogram where the original voiced speech signal (bottom left) is converted into a whispered speech signal (i.e., the pitch is removed).}
     \label{fig:qualitative_examples}
\end{figure*}

\subsection{Qualitative results}

In Figure~\ref{fig:qualitative_examples}, we illustrate the ability of the proposed method to modify $f_0$ and the formant frequencies in an accurate and independent manner. The top-left spectrogram contains five segments of equal length. The first segment corresponds to the original spectrogram of the steady vowel /a/ uttered by a female speaker. In the following segments, we vary successively each individual factor $f_i$, for $i=0$ to $3$, as indicated by the black lines in the figure. Variations of $f_0$ modify the harmonic structure of the signal while keeping the formant structure unaltered. Variations of $f_i$, $i \in \{1,2,3\}$, modify the formant frequencies, as indicated by the color map, while keeping $f_0$ unaltered. 

The top-right spectrogram in Figure~\ref{fig:qualitative_examples} was generated by using the conditional prior in \Eqref{conditional_prior} (generalized to conditioning on multiple factors). We can see that the characteristics of the generated speech spectrogram match well with the input trajectories represented by the lines in the figure.

In the second row of Figure~\ref{fig:qualitative_examples}, from left to right we show the original spectrogram of a speech signal uttered by a female speaker (left), the transformed spectrogram where $f_0$ is set constant over time (middle), and the transformed spectrogram where the pitch has been removed (i.e., the original voiced speech signal is converted into a whispered speech signal) (right). This last spectrogram is obtained by subtracting to $\mathbf{z}$ its projection onto the latent subspace corresponding to $f_0$ (i.e., by considering only the two first terms in the right-hand side of \Eqref{z_tilde}). This results in a spectrogram where the harmonic component is neutralized, while preserving the original formant structure. This is remarkable considering that the VAE was not trained on whispered speech signals, and it further confirms that the proposed method dissociates the source and the filter contributions in the VAE latent space. 

Audio examples and additional examples of generated and transformed speech spectrograms can be found online.\footnote{ \href{https://samsad35.github.io/site-sfvae/}{https://samsad35.github.io/site-sfvae/}} In \ref{app:viz}, we provide plots of trajectories in the learned latent subspaces, illustrating that, according to each factor, the proximity of two speech spectra is preserved in the corresponding latent subspace.

\subsection{Quantitative results}

\begin{figure}[t]
    \centering
    \includegraphics[width=.5\linewidth]{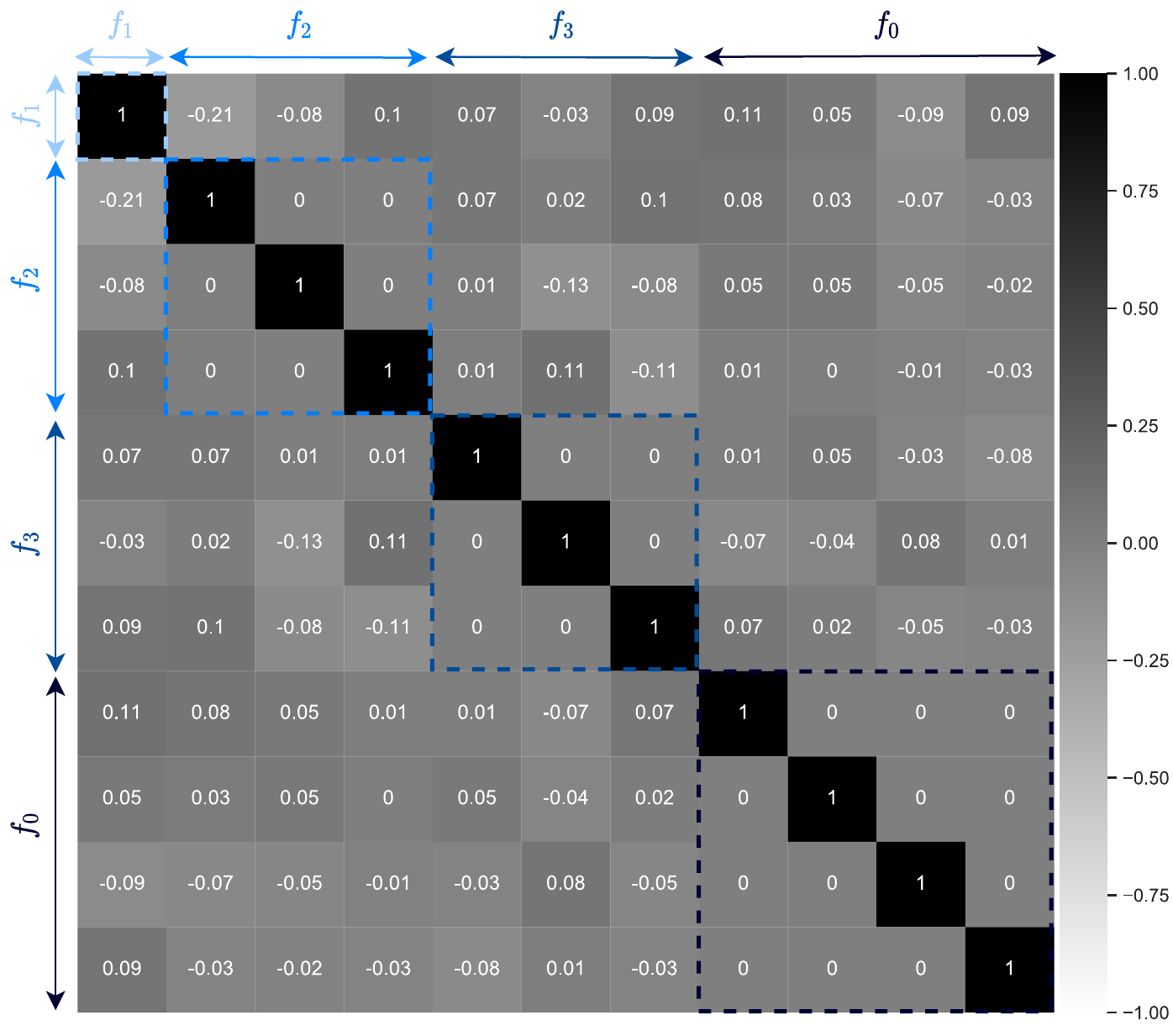}
    \caption{Correlation matrix of the learned latent subspaces basis vectors.}%
    \label{fig:orthogonality}
\end{figure}

\subsubsection{Orthogonality of the latent subspaces}
\label{subsec:Orthogonality}
In this experiment, we quantitatively evaluate the proposed method in terms of disentanglement of the learned source-filter latent representation. Following the discussion in Section~\ref{subsec:orthogonality}, we compute the dot product between all pairs of unit vectors in the matrices $\mathbf{U}_i \in \mathbb{R}^{L \times M_i}$, $i \in \{0,1,2, 3\}$. Figure~\ref{fig:orthogonality} shows that the resulting correlation matrix is mainly diagonal. Except for a correlation value of $-0.21$ across $f_1$ and the first component of $f_2$, all other values are below $0.13$ (in absolute value), confirming the orthogonality of the learned subspaces and thus the disentanglement of the learned source-filter representation of speech. 
The methodology presented in Section~\ref{subsec:learning_latent_subspaces} was applied to try to identify orthogonal source-filter subspaces directly in the space of speech power spectra and Mel-frequency cepstral coefficients (MFCCs). The results (see \ref{app:correlation_mfcc_spec}) show that an organization into orthogonal source-filter subspaces does not exist when working on such raw signal representations. Indeed, there exist strong correlations between the learned subspaces. This confirms that the disentanglement in terms of source-filter factors is achieved during the unsupervised learning of the VAE model. We remind that in the proposed method, the synthetic labeled data are only used to identify the disentangled subspaces of the learned representation, after the VAE unsupervised training.

\subsubsection{Pitch and formant transformations} 
\label{Experimental}

In this experiment, we quantitatively evaluate the performance of the proposed method regarding the modification of $f_0$ and the formant frequencies in speech signals (see Section~\ref{subsec:control_factor}).

\paragraph{Experimental set-up}  We use a corpus of 12 English vowels uttered by 50 male and 50 female speakers \citep{hillenbrand1995acoustic}, which is labeled with the value of $f_0$ and the formant frequencies. We also use the TIMIT dataset \citep{TIMIT}, a corpus of phonemically and lexically transcribed speech of American English speakers of different sexes and dialects. We used the test corpus containing 1680 utterances. Because we are interested in studying the interaction between modifications of $f_0$ and the formant frequencies, we only evaluate the method on the voiced phonemes (40 phonemes over a total of 52), which are identified using the annotations. We transform each test signal in the English vowels and TIMIT datasets by varying one single factor $f_i$ at a time, for $i \in \{0,1,2,3\}$, according to the ranges and step sizes given in Table~\ref{tab:test_dataset}. For instance, when performing transformations of $f_0$, for each test signal in the English vowels dataset, we vary linearly the target $f_0$ value between 100 and 300 Hz, with a step size of 1 Hz, thus resulting in 200 transformations.

\begin{table}
\centering
\begin{tabular}{c c c c|cc}
\multirow{2}{*}{Factor} & Min & Max & \multicolumn{2}{c}{Step size (in Hz)} & Relative \\
\noalign{\vskip 1ex} \cline{4-5} \noalign{\vskip 1ex}
& \multicolumn{2}{c}{(in Hz)} & English vowels & TIMIT &  variation (\%) \\ 
\noalign{\vskip 1ex} \hline \noalign{\vskip 1ex}
$f_0$ & 100 & 300 & 1 & 10 & $\pm$50\\
$f_1$ & 300 & 900 & 10 & 50 & $\pm$50\\
$f_2$ & 1100 & 2700 & 20 & 100 & $\pm$42\\
$f_3$ & 2200 & 3200 & 20 & 50 & $\pm$18
\end{tabular}
\caption{Variation range (min and max values) and step size used for the transformation of each test signal in the English vowels and TIMIT datasets, for each factor of variation $f_i$, $i \in \{0,1,2,3\}$. The last column indicates by how much a factor varies relative to the center value of its variation range. Its entries are computed as $\pm$ (max $-$ min)/(max + min)$\times 100 \%$.}
\label{tab:test_dataset}
\end{table}

\paragraph{Metrics} For the modification of each factor $f_i$, we measure the performance regarding three aspects: First, in terms of  \textit{accuracy} by comparing the target value for the factor (see \Eqref{z_tilde}) and its estimation computed from the modified output speech signal. Second, in terms of \textit{disentanglement}, by comparing the values of $f_j$ for $j \neq i$, before and after modification of the factor $f_i$. Third, in terms of speech \textit{naturalness} of the transformed signal. 

Accuracy and disentanglement are measured in terms of relative magnitude error (in percent, the lower the better). For a given factor $f_i$, it is defined by $\delta f_{i} = 100\% \times |\hat{y} - y|/y$ where $y$ is the target value of $f_i$ and $\hat{y}$ its estimation from the output transformed signal. Let us take the example of a modification of $f_0$: $\delta f_{0}$ measures the accuracy of the transformation on $f_0$ while $\delta f_{1}$, $\delta f_{2}$ and $\delta f_{3}$ are used to assess if the other factors of variation $f_1$, $f_2$ and $f_3$ remained unchanged after modifying $f_0$. We use CREPE \citep{kim2018crepe} to estimate $f_0$ and Parselmouth \citep{parselmouth}, which is based on PRAAT \citep{praat}, to estimate the formant frequencies. 
Regarding speech naturalness, we use the NISQA objective measure \citep{mittag2020deep}. This metric (the higher the better) was developed in the context of speech transformation algorithms and it was shown to highly correlate with subjective mean opinion scores (MOS) (i.e., human ratings). As a reference, the NISQA score on the original dataset of English vowels (i.e., without any processing) is equal to 2.60~$\pm$~0.53.

TIMIT is phonemically richer than the English vowels dataset, however, it is not labeled with $f_0$ and formant frequency values. Therefore, we do not have the ground truth values which makes the evaluation in terms of disentanglement more difficult than with the English vowels labeled dataset. Instead of the ground truth, we use the formant frequencies and $f_0$ values computed on the original speech utterances (i.e., before transformation). This makes the evaluation on TIMIT less reliable than on the English vowels dataset, but it allows us to test the methods on a larger variety of phonemes.

\paragraph{Methods} We compare the proposed approach with several methods from the literature: (i) TD-PSOLA \citep{moulines1990pitch} performs $f_0$ modification through the decomposition of the signal into pitch-synchronized overlapping frames. (ii) WORLD \citep{morise2016world} is a vocoder also used for $f_0$ modification. It decomposes the speech signal into three components characterizing $f_0$, the aperiodicity, and the spectral envelope. (iii) The method proposed by \cite{hsu2017learning} (here referred to as ``VAE baseline'') consists in applying translations directly in the latent space of the VAE. Unlike the proposed approach, this method requires predefined latent attribute representations $\boldsymbol{\mu}_{\text{src}}$ and $\boldsymbol{\mu}_{\text{trgt}}$ associated with the source and target values of the factor to be modified, respectively. In particular, computing $\boldsymbol{\mu}_{\text{src}}$ requires analyzing the input speech signal, for instance, to estimate $f_0$, which is not the case for the proposed method. The source and target latent attribute representations are then used to perform the translation $\tilde{\mathbf{z}} = \mathbf{z} - \boldsymbol{\mu}_{\text{src}} + \boldsymbol{\mu}_{\text{trgt}}$, where $\mathbf{z}$ and $\tilde{\mathbf{z}}$ are respectively the original and modified latent vectors. To ensure a fair comparison, we build dictionaries of predefined latent attribute representations using the same artificially-generated speech data that were used in the training stage of the proposed method. All the methods we compare with require a pre-processing of the input speech signal to compute the input trajectory of the factor to be modified, which is not the case of the proposed method.

\begin{table*}
\resizebox{\linewidth}{!}{ 
\begin{tabular}{lllllll|lllll}
& & \multicolumn{5}{c}{English vowels dataset} & \multicolumn{5}{c}{TIMIT dataset} \\

\noalign{\vskip 1ex} \cline{3-7} \cline{8-12} \noalign{\vskip 1ex}

Factor & Method & { NISQA  \tiny{($\uparrow$)}} & {$\delta f_0$ {\tiny(\%, $\downarrow$)}} & {$\delta f_1$ {\tiny(\%, $\downarrow$)}} & {$\delta f_2$ {\tiny(\%, $\downarrow$)}} &
{$\delta f_3$ {\tiny(\%, $\downarrow$)}} & { NISQA  \tiny{($\uparrow$)}} & {$\delta f_0$ {\tiny(\%, $\downarrow$)}} & {$\delta f_1$ {\tiny(\%, $\downarrow$)}} & {$\delta f_2$ {\tiny(\%, $\downarrow$)}} &
{$\delta f_3$ {\tiny(\%, $\downarrow$)}} \\

\noalign{\vskip 1ex} \hline \noalign{\vskip 1ex}

$f_0$   & TD-PSOLA   & 2.32 {\tiny$\pm$ 0.55} &  3.8 {\tiny$\pm$ 2.5} & 6.3 {\tiny$\pm$ 2.8} & 3.7 {\tiny$\pm$ 0.9} & 2.1 {\tiny$\pm$ 0.5}  & 2.36 {\tiny$\pm$ 0.50} &  2.4 {\tiny$\pm$ 1.9} & 7.9 {\tiny$\pm$ 0.6} & 4.5 {\tiny$\pm$ 0.3} & 3.9 {\tiny$\pm$ 0.2}  \\
& WORLD & 2.49 {\tiny$\pm$ 0.60}  & 4.5 {\tiny$\pm$ 0.6} & 3.7 {\tiny$\pm$ 1.8} & 2.3 {\tiny$\pm$ 0.7} & 1.2 {\tiny$\pm$ 0.2}  & 2.45 {\tiny$\pm$ 0.47}  & 0.3 {\tiny$\pm$ 0.1} & 7.1 {\tiny$\pm$ 1.2} & 6.2 {\tiny$\pm$ 0.4} & 4.2 {\tiny$\pm$ 0.2}  \\
& VAE baseline  & 1.94 {\tiny$\pm$  0.43} &  6.21 {\tiny$\pm$2.8}& 10.4 {\tiny$\pm$ 2.4} & 6.2 {\tiny$\pm$ 0.9} & 4.5 {\tiny$\pm$ 0.2} & 1.59 {\tiny$\pm$  0.43} &  16.1 {\tiny$\pm$6.3}& 17.0 {\tiny$\pm$ 3.0} & 12.1{\tiny$\pm$ 0.2} & 10.9 {\tiny$\pm$ 1.3} \\
& Proposed & 2.08 {\tiny$\pm$ 0.48} & 0.8 {\tiny$\pm$ 0.2} & 7.2 {\tiny$\pm$ 1.3} & 3.6 {\tiny$\pm$ 1.2} & 3.8 {\tiny$\pm$ 0.3} & 2.28 {\tiny$\pm$ 0.57} & 0.8 {\tiny$\pm$ 0.6} & 9.1 {\tiny$\pm$ 1.1} & 8.3 {\tiny$\pm$ 0.9} & 6.0 {\tiny$\pm$ 1.8}\\

\noalign{\vskip 1ex} \hline \noalign{\vskip 1ex}

$f_1$   & VAE baseline & 1.84 {\tiny$\pm$  0.5} & 11.3 {\tiny$\pm$ 4.2}         & 15.1 {\tiny$\pm$ 3.5} & 6.0 {\tiny$\pm$ 1.2}  & 4.2 {\tiny$\pm$ 0.4} & 1.42 {\tiny$\pm$  0.34} & 10.1 {\tiny$\pm$ 2.8}         & 16.4 {\tiny$\pm$ 1.4} & 12.4 {\tiny$\pm$ 0.9}  & 11.2 {\tiny$\pm$ 2.6}\\
& Proposed & 1.85 {\tiny$\pm$ 0.4 } & 6.0 {\tiny$\pm$ 1.6} & 8.4 {\tiny$\pm$ 3.2} & 5.7 {\tiny$\pm$  0.4} & 4.4 {\tiny$\pm$ 0.3} & 1.66 {\tiny$\pm$ 0.31 } & 7.1 {\tiny$\pm$ 3.6} & 9.2 {\tiny$\pm$ 0.8} & 9.0 {\tiny$\pm$  1.3} & 7.8 {\tiny$\pm$ 1.1} \\

\noalign{\vskip 1ex} \hline \noalign{\vskip 1ex}

$f_2$   & VAE baseline & 2.01 {\tiny$\pm$  0.4}  & 19.5 {\tiny$\pm$ 3.2}& 10.7 {\tiny $\pm$ 0.5} & 10.9 {\tiny$\pm$ 1.9} & 5.8 {\tiny$\pm$ 0.6}  & 1.46 {\tiny$\pm$  0.30}  & 19.3 {\tiny$\pm$ 5.0}& 16.4 {\tiny $\pm$ 0.8} & 20.3 {\tiny$\pm$ 6.3} & 11.5 {\tiny$\pm$ 0.5}\\
& Proposed & 2.03 {\tiny$\pm$ 0.43 }  & 8.5 {\tiny$\pm$ 1.1} & 8.7 {\tiny$\pm$ 1.1} & 6.2 {\tiny$\pm$ 1.5} & 5.8 {\tiny$\pm$ 0.2} & 1.49 {\tiny$\pm$ 0.30 }  & 9.1 {\tiny$\pm$ 2.2} & 8.3 {\tiny$\pm$ 1.3} & 4.3 {\tiny$\pm$ 1.3} & 8.1 {\tiny$\pm$ 0.2}\\

\noalign{\vskip 1ex} \hline \noalign{\vskip 1ex}

$f_3$   & VAE baseline & 1.82 {\tiny$\pm$  0.14}  & 27.0 {\tiny$\pm$ 1.5}& 13.0 {\tiny $\pm$ 1.3} & 12.0 {\tiny$\pm$ 1.8} & 7.3 {\tiny$\pm$ 1.5} & 1.40 {\tiny$\pm$  0.48}  & 20.4 {\tiny$\pm$ 1.0}& 17.4 {\tiny $\pm$ 0.2} & 14.4 {\tiny$\pm$ 0.2} & 11.7 {\tiny$\pm$ 2.3}\\
& Proposed & 1.94 {\tiny$\pm$ 0.48 }  & 8.3 {\tiny$\pm$ 1.0} & 8.6 {\tiny$\pm$ 0.7} & 4.9 {\tiny$\pm$ 0.9} & 2.0 {\tiny$\pm$ 0.4} & 1.48 {\tiny$\pm$ 0.42 }  & 8.5 {\tiny$\pm$ 1.9} & 8.7 {\tiny$\pm$ 0.9} & 5.7 {\tiny$\pm$ 2.1} & 2.5 {\tiny$\pm$ 1.8}\\

\end{tabular}
}
\caption{Performance (mean and standard deviation) for the transformation of $f_0$ and the formant frequencies ($f_1$, $f_2$ and $f_3$) on the English vowel and TIMIT datasets.}
\label{tab:all_results}
\end{table*}

\paragraph{Discussion} 
\label{sec:discussion}
Experimental results (mean and standard deviation) are shown in Table~\ref{tab:all_results}. Compared to the VAE baseline, the proposed method obtains better performance in terms of accuracy, disentanglement, and naturalness, for both test datasets.
These results confirm the effectiveness of performing the transformations in the learned latent subspaces and not directly in the latent space, as well as the advantage of using regression models instead of predefined latent attribute representations. 
To analyze the disentanglement results, when performing the transformation for a given factor $f_i$, one must compare the movement of other factors $\{f_j, j \neq i\}$ relative to their fixed targets (as given by the metrics $\{\delta f_j, j \neq i\}$) to the movement of the factor $f_i$ that is varied relative to the center value of its variation range (as given in the last column of Table~\ref{tab:test_dataset}). For instance, when $f_0$ is varied in the experiments on the English vowels dataset, Table~\ref{tab:all_results} shows that the proposed method makes the other factors move from their fixed targets by $7.2 \%$, $3.6 \%$, and $3.8 \%$ for $f_1$, $f_2$, and $f_3$ respectively. These values are much smaller than the relative variation of the factor $f_0$ itself, as indicated in the last column of Table~\ref{tab:test_dataset} is equal to $50\%$. We can thus conclude that modifying $f_0$ has little effect on the other factors. Similar conclusions can be drawn by analyzing the disentanglement results for the variation of other factors, confirming the disentanglement of the learned representation.
Regarding $f_0$ transformation, WORLD obtains the best performance in terms of disentanglement, which is because the source and filter contributions are decoupled in the architecture of the vocoder. In terms of naturalness, WORLD and then TD-PSOLA obtain the best performance. This may be explained by the fact that these methods operate directly in the time domain, therefore they do not suffer from phase reconstruction artifacts, unlike the proposed and VAE baseline methods. Naturalness is indeed greatly affected by phase reconstruction artifacts, even from an unaltered speech spectrogram (i.e., without transformation). Phase reconstruction in a multi-speaker setting is still an open problem in speech processing. We want to emphasize that the objective of this study is not to compete with traditional signal processing methods such as TD-PSOLA and WORLD. It is rather to advance on the understanding of deep generative modeling of speech signals and to compare honestly with highly-specialized traditional systems. TD-PSOLA and WORLD exploit signal models specifically designed for the task at hand, which for instance prevents them to be used for modifying formant frequencies. In contrast, the proposed method is fully based on learning and the same methodology applies to modifying $f_0$ or the formant frequencies.

\subsubsection{Robustness with respect to the VAE training dataset}
\label{additional_results_otherdataset}

In this Section, we investigate the robustness of the proposed method with respect to different datasets used to train the VAE model. We considered three training datasets in addition to the WSJ0 dataset used in the previous experiments: (i) the SIWIS French speech synthesis dataset \citep{honnet2017siwis}, which contains more than ten hours of French speech recordings; (ii) the Toronto emotional speech (TESS) dataset \citep{dupuis2010toronto}, which contains 2,800 utterances spoken by two actresses using different emotions (anger, disgust, fear, happiness, pleasant surprise, sadness, and neutral); and (iii) the LJspeech dataset \citep{ljspeech17}, which contains 13,100 short audio clips of a single speaker reading passages from 7 non-fiction books. The artificially-generated speech dataset used for learning the latent subspaces and the regression models remain the same. 

Table~\ref{tab:robustness_results} presents the results for the modification of $f_0$ only, applied to the English vowels dataset. It can be seen in Table~\ref{tab:robustness_results} that the performance remains quite stable with different VAE training datasets. WSJ0 is the largest dataset and therefore leads to the best performance. Interestingly, the results obtained with the SIWIS dataset of French speech signals remain satisfactory, even if there is a mismatch between the training (French) and testing (English) datasets.

\begin{table}
\centering
\begin{tabular}{llllllll}
Dataset & NISQA  \tiny{($\uparrow$)} & $\delta f_0$ {\tiny(\%, $\downarrow$)} & $\delta f_1$ {\tiny(\%, $\downarrow$)} & $\delta f_2$ {\tiny(\%, $\downarrow$)} &$\delta f_3$ {\tiny(\%, $\downarrow$)} 
\\ \hline \\
WSJ &  2.08 {\tiny$\pm$ 0.48} & 0.8 {\tiny$\pm$ 0.2} & 7.2 {\tiny$\pm$ 1.3} & 3.6 {\tiny$\pm$ 1.2} & 3.8 {\tiny$\pm$ 0.3}\\
SIWIS &  1.93 {\tiny$\pm$ 0.43} & 1.2 {\tiny$\pm$ 0.5} & 10.0 {\tiny$\pm$ 4.2} & 8.3 {\tiny$\pm$ 1.1} & 14.0 {\tiny$\pm$ 0.2}\\
TESS &  1.98 {\tiny$\pm$ 0.50} &  2.7 {\tiny$\pm$ 2.3} & 9.3 {\tiny$\pm$ 3.5} & 9.0 {\tiny$\pm$ 0.8} & 7.0 {\tiny$\pm$ 0.2}\\
LJspeech  & 1.96 {\tiny$\pm$ 0.40} & 1.2 {\tiny$\pm$ 0.6} & 9.3 {\tiny$\pm$ 1.2} & 5.6 {\tiny$\pm$ 0.6} & 4.6 {\tiny$\pm$ 0.1}\\
\end{tabular}
\caption{Performance (mean and standard deviation) of $f_0$ transformation with the proposed method, on the English vowels test dataset, using different training datasets for the unsupervised VAE model.}
\label{tab:robustness_results}
\end{table}

\begin{figure*}    
     \centering
     \begin{subfigure}{0.32\textwidth}
         \centering
         \includegraphics[width=1.0\textwidth]{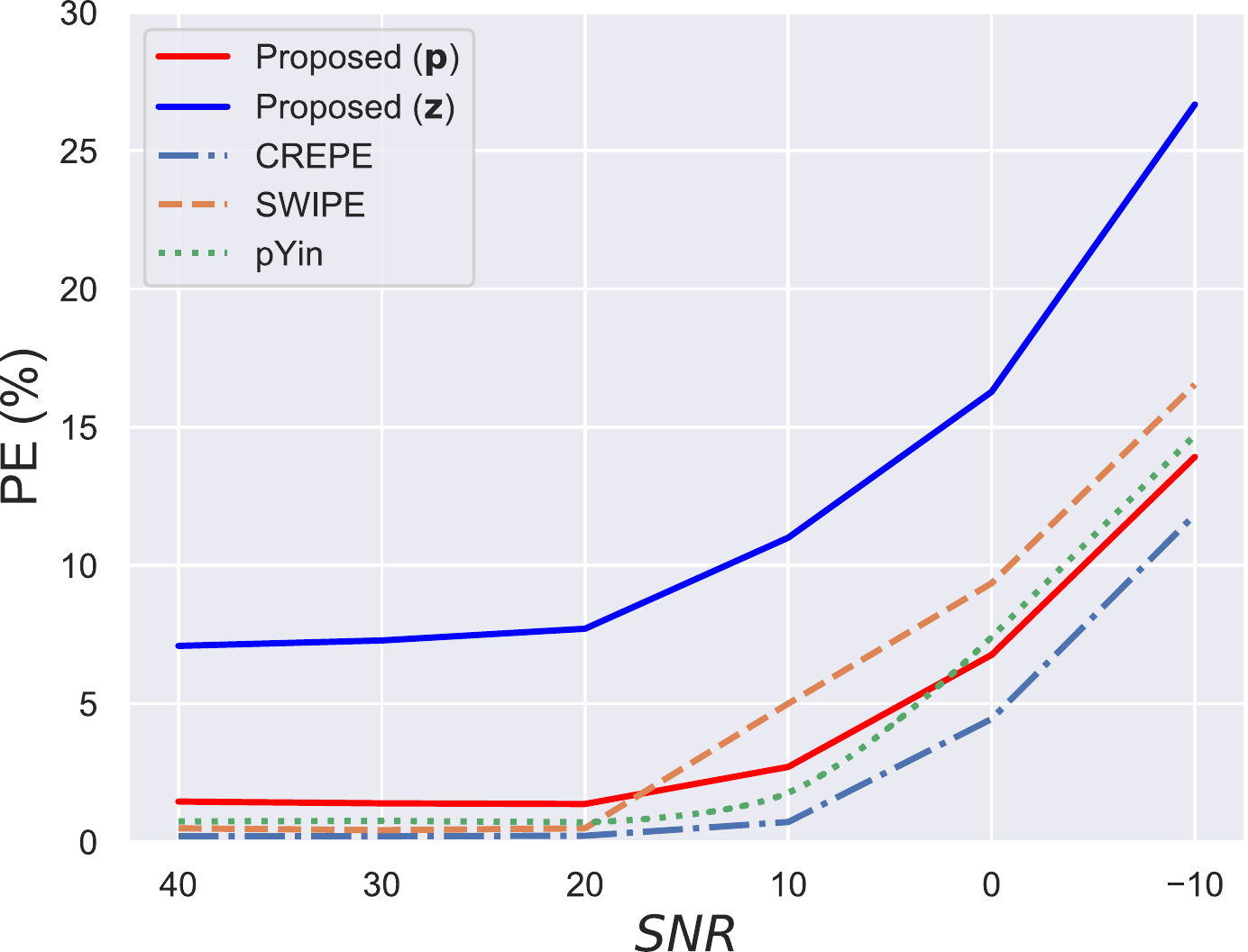}
         \caption{$\lambda = 20\%$}
         \label{fig:pitch-detection-20}
     \end{subfigure}
     \begin{subfigure}{0.32\textwidth}
         \centering
         \includegraphics[width=1.0\textwidth]{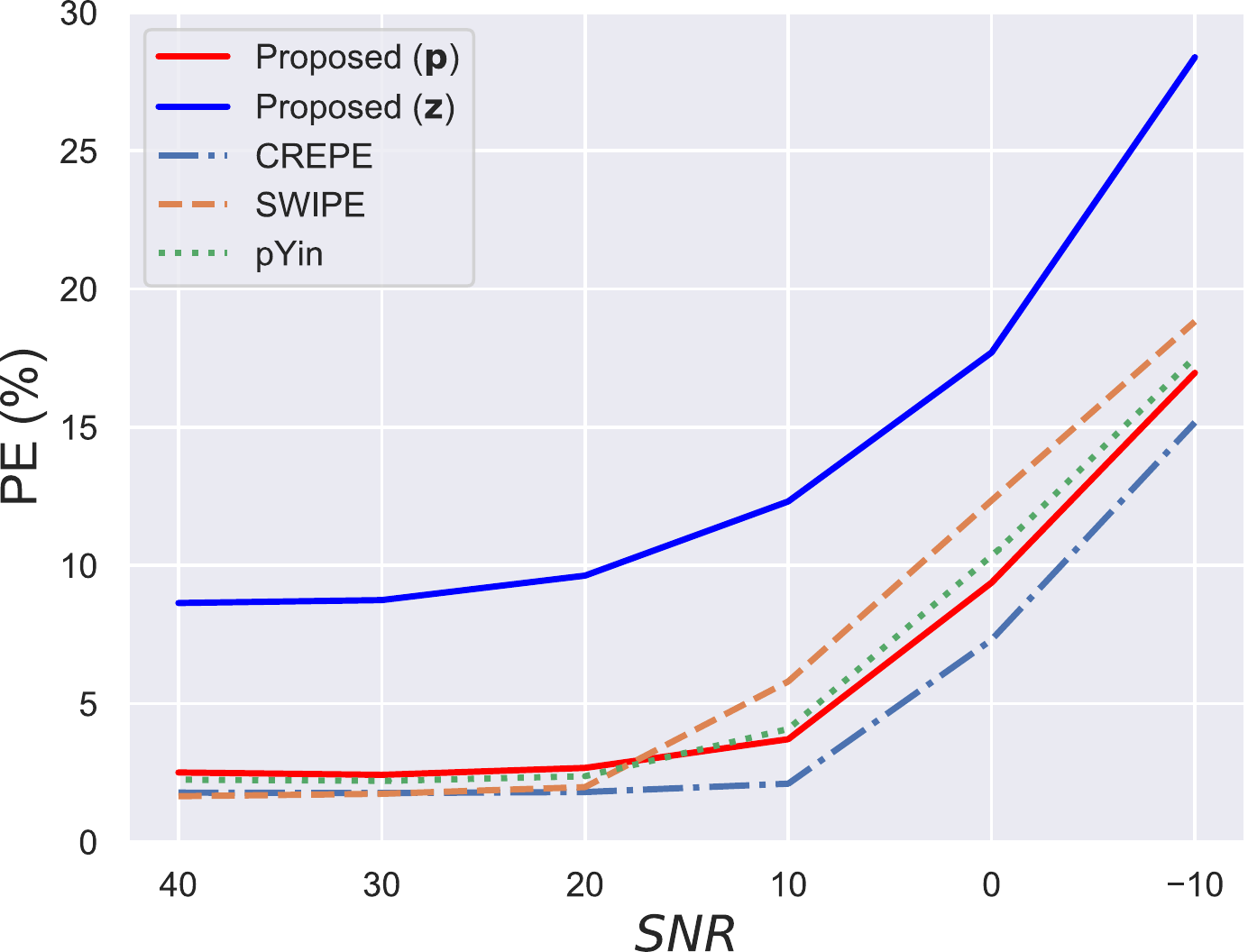}
         \caption{$\lambda = 10\%$}
         \label{fig:pitch-detection-10}
     \end{subfigure}
     \begin{subfigure}{0.32\textwidth}
         \centering
         \includegraphics[width=1.0\textwidth]{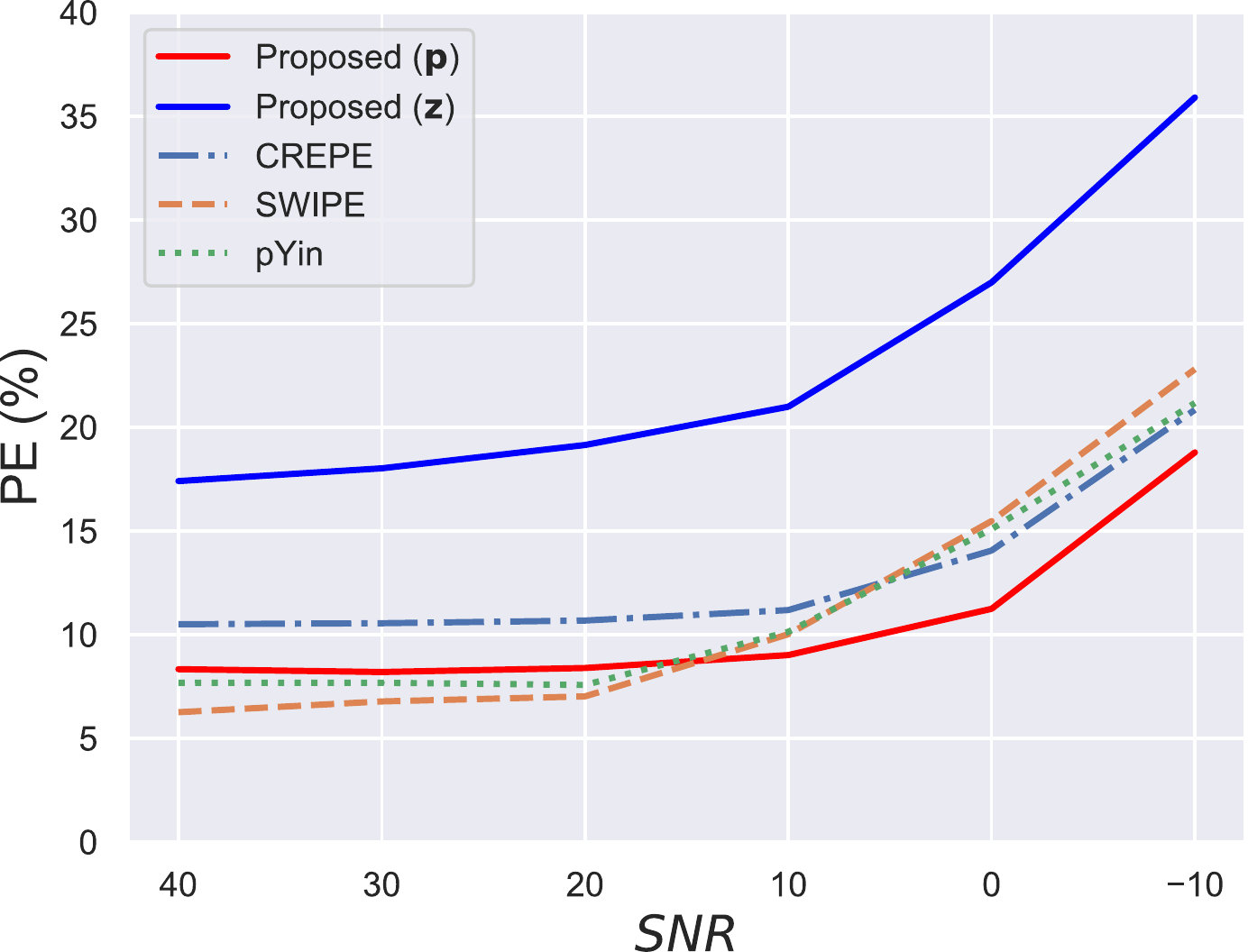}
         \caption{$\lambda = 1\%$}
         \label{fig:pitch-detection-01}
     \end{subfigure}
     \caption{Results of the $f_0$ tracking experiment: Pitch error (PE, in \%) as a function of the SNR (in dB), for different values of the threshold $\lambda$ (in \%).  
     ``Proposed($\mathbf{p}$)" and ``Proposed($\mathbf{z}$)" denote the proposed approach for $f_0$ estimation using the projection of $\mathbf{z}$ into the learned subspace of the pitch and using $\mathbf{z}$ directly without the projection, respectively. }
     \label{fig:pitch-detection}
\end{figure*}

\subsubsection{Robust $f_0$ estimation}
\label{pitch-detection}

\paragraph{Experimental set-up}
For the experiments on fundamental frequency estimation, we used the Pitch Tracking Database from Graz University of Technology (PTDB-TUG) \citet{pirker2012pitch}. This dataset provides microphone and laryngograph signals from 20 native English speakers. The ground truth $f_0$ is extracted from the laryngograph signals. To evaluate the robustness of the $f_0$ estimation methods, we corrupted each clean speech test signal by adding real-world cafeteria noise extracted from the DEMAND dataset \citep{thiemann2013demand}, where the signal-to-noise ratio (SNR) is varied from $-10$ to $40$~dB. In this experiment, for both the proposed and reference methods, we only considered the problem of estimating $f_0$ given the ground-truth voicing labels, i.e., we did not consider the problem of detecting which frames are voiced.

\paragraph{Metric} 
Following \citet{rabiner1976comparative}, the performance is measured in terms of pitch error (PE) defined as the proportion of frames considered as voiced by both the estimation algorithm and the ground truth for which the relative $f_0$ error is higher than a certain threshold $\lambda$ in \%:
\begin{equation}
    PE = \frac{N_{est}}{N_{v}}~\times~100 \%,
\end{equation}
where $N_{v}$ is the number of voiced frames and $N_{est}$ is the number of the frames for which $| \hat{y} - y | / y \ge \lambda$ with $\hat{y}$ and $y$ the estimated and ground-truth $f_0$ values, respectively. The parameter $\lambda$ can be interpreted as the tolerance for which the $f_0$ prediction is considered as correct. 

\paragraph{Methods}
We compared the proposed approach described in Section~\ref{subsec:analyse_factor} with several state-of-the-art methods from the literature:  
(i) pYin \citep{mauch2014pyin} improvement of the Yin-based autocorrelation algorithm in the time domain using probabilistic threshold distribution. We used the one implemented in the Librosa package \citep{mcfee2015librosa}; 
(ii) SWIPE \citep{camacho2008sawtooth} is a method for pitch detection based on the autocorrelation of the speech signal in the frequency domain. We used the one implemented in the pysptk toolkit \citep{yamamoto2019r9y9}; 
(iii) CREPE \citep{kim2018crepe} is a supervised method based on CNN already used in Section~\ref{Experimental}. We used the one implemented in the torchcrepe github of \cite{morrison2020torchcrepe}. 
To confirm the interest of estimating the $f_0$ in the corresponding latent subspace and not directly in the VAE latent space, we also apply the proposed method described in \ref{subsec:analyse_factor} by directly working with $q_\phi(\mathbf{z} | \mathbf{x})$ instead of $q_\phi(\mathbf{p} | \mathbf{x})$ to compute the KL divergence in \eqref{KL_f0_est}. This aims to show that projecting the speech signal in the latent subspace encoding $f_0$ indeed helps discarding information that is not related to this factor and thus improves the estimation accuracy.

\paragraph{Discussion}
The results are shown in Figure~\ref{fig:pitch-detection}. Figures~\ref{fig:pitch-detection-20}, \ref{fig:pitch-detection-10}, and~\ref{fig:pitch-detection-01} display the PE metric (\%) as a function of the SNR for $\lambda$ equal to 20\%, 10\%, and 1\%, respectively.  
The proposed pitch estimation method based on the projection of the latent variable $\mathbf{z}$ on the $f_0$ subspace outperforms the same approach using the latent variable $\mathbf{z}$ directly. This result confirms the fact that the $f_0$ subspace is less sensitive to formant variations (invariance property), and projecting $\mathbf{z}$ into this subspace globally preserves well the pitch information, which makes the detection more robust to variations that are independent of $f_0$. 
For large $\lambda$ values (i.e., high precision tolerance), the proposed method gives results that are competitive with the state-of-the-art methods for moderate noise levels (high SNRs) and that outperform SWIPE for a high level of noise (low SNRs).
For small $\lambda$ values (i.e., low precision tolerance), the proposed approach shows the best trade-off between robustness to noise and precision. Indeed, the proposed method is about $2$\% PE below CREPE, consistently over the whole SNR range. It also outperforms SWIPE and pYin below $14$dB SNR.

\section{Conclusion}
\label{sec:conclusion}

The source-filter model of speech production is a fundamental concept in speech processing. In this work, using only a few seconds of artificially generated labeled speech data, we showed that the fundamental frequency and formant frequencies are encoded in orthogonal latent subspaces of an unsupervised VAE. We proposed to exploit this disentangled source-filter latent representation for the transformation and analysis of speech spectrograms. Using a regression model trained on the artificially-generated labeled speech data, we proposed a method to control $f_0$ and the formant frequencies in speech spectrograms by applying affine transformations in the learned latent subspaces. We also proposed to exploit the projection of the speech signal in the latent subspace associated with $f_0$ to robustly estimate the latter from speech signals corrupted by noise. Even if the identification of the latent source-filter subspaces, the learning of the regression models, and the $f_0$ estimation method were designed using a very limited set of artificially-generated signals, we showed experimentally that the speech transformations and analysis are effective on natural signals.
To the best of our knowledge, this is the first approach that, with a single methodology, is able to extract, identify and control the source and filter low-level speech attributes within a VAE latent space. This is an important step towards a better understanding of deep latent-variable generative modeling of speech signals.

Future work includes improving the quality of the generated speech waveforms, by addressing the phase reconstruction issue or by directly modeling the speech waveform \citep{caillon2021rave}. It also includes extending the proposed method to dynamical VAEs \citep{girin2020dynamical}, to hierarchical latent spaces \citep{vahdat2020nvae}, and to audio-visual speech processing. The proposed model could also be applied to pitch-informed speech enhancement. Indeed, several recent weakly-supervised speech enhancement methods consist in estimating the VAE latent representation of a clean speech signal given a noisy speech signal \citep{bando2018statistical,leglaive_MLSP18, BayesianMVAE, Leglaive_ICASSP2019a, Leglaive_ICASSP2019b, parienteInterspeech19, leglaive2020recurrent, richter2020speech, carbajal2021guided, fang2021variational}. Using the proposed conditional deep generative speech model, this estimation could be constrained given the $f_0$ contour of the signal.

\appendix
\section{Solution to the latent subspace learning problem}
\label{app:PCA}

In this Appendix, we show that the solution to the optimization problem \plaineqref{pca_opt_prb_1} is given by the principal eigenvectors of $\mathbf{S}_\phi(\mathcal{D}_i)$ in \Eqref{S_Di}. Without loss of generality, we formulate the problem for a centered version of the latent data:
\begin{equation}
    \mathbf{z} \leftarrow \mathbf{z} - \boldsymbol{\mu}_\phi(\mathcal{D}_i),
\end{equation}
where $\boldsymbol{\mu}_\phi(\mathcal{D}_i)$ is defined in \Eqref{mu_Di}. This centering also affects the inference model originally defined in \Eqref{vae_inf_dist}, as follows:
\begin{equation}
    q_\phi(\mathbf{z} | \mathbf{x}) = \mathcal{N}\left(\mathbf{z}; \boldsymbol{\mu}_{\phi}(\mathbf{x}) - \boldsymbol{\mu}_\phi(\mathcal{D}_i), \diag\{\mathbf{v}_{\phi}(\mathbf{x}) \right).
    \label{inf_model_center}
\end{equation}
Using \Eqref{agg_posterior}, the fact that $\mathbf{U}^\top \mathbf{U} = \mathbf{I}$, and \Eqref{inf_model_center}, the cost function in the optimization problem \plaineqref{pca_opt_prb_1} can be rewritten as follows:
\begin{align}
    \mathbb{E}_{\hat{q}_\phi^{(i)}(\mathbf{z})} \left[ \left\lVert \mathbf{z} - \mathbf{U} \mathbf{U}^\top \mathbf{z} \right\rVert_2^2 \right] & = \frac{1}{\# \mathcal{D}_i} \sum_{\mathbf{x}_n \in \mathcal{D}_i} \mathbb{E}_{q_\phi(\mathbf{z} | \mathbf{x}_n)}\left[ \left\lVert \mathbf{z} - \mathbf{U} \mathbf{U}^\top \mathbf{z} \right\rVert_2^2 \right] \nonumber \\
    & = \tr\left\{ (\mathbf{I} - \mathbf{U} \mathbf{U}^\top) \frac{1}{\# \mathcal{D}_i} \sum_{\mathbf{x}_n \in \mathcal{D}_i} \mathbb{E}_{q_\phi(\mathbf{z} | \mathbf{x}_n)}[ \mathbf{z} \mathbf{z}^\top] \right\} \nonumber \\
    & = \tr\left\{ (\mathbf{I} - \mathbf{U} \mathbf{U}^\top) \mathbf{S}_\phi(\mathcal{D}_i) \right\},
\end{align}
where $\mathbf{S}_\phi(\mathcal{D}_i)$ is defined in \Eqref{S_Di}. From this last equality, we see that the optimization problem~\plaineqref{pca_opt_prb_1} is equivalent to
\begin{equation}
    \max_{\mathbf{U} \in \mathbb{R}^{L \times M_i}} \tr\left\{ \mathbf{U}^\top \mathbf{S}_\phi(\mathcal{D}_i) \mathbf{U} \right\}, \qquad s.t.\,\, \mathbf{U}^\top \mathbf{U} = \mathbf{I}.
\end{equation}
Very similarly to PCA \citep{pearson1901liii}, the solution is given by the $M_i$ dominant eigenvectors of $\mathbf{S}_\phi(\mathcal{D}_i)$ (i.e., associated to the $M_i$ largest eigenvalues) \citep[Section 12.1]{Bishop06}.

\section{Algorithms}
\label{app:algorithms}

\algrenewcommand{\algorithmiccomment}[1]{\hskip.1em$\#$ \textit{#1}}

\begin{algorithm}[H] 
\caption{Learning of the regression model.}
\label{alg:regression}
\begin{algorithmic}[1]

\Require{}
\Statex{$\bullet$ $\mathcal{D}_i = \{\mathbf{x}_n \in \mathbb{R}_+^D, y_n \in \mathbb{R}_+\}_{n=1}^N$: Dataset of artificially-generated speech power spectra $\mathbf{x}_n$ and the corresponding values $y_n$ for the factor $f_i$}
\Statex{$\bullet$ $\mathbf{U}_i \in \mathbb{R}^{L \times M_i}$: Latent subspace matrix associated with $f_i$ obtained by solving \plaineqref{pca_opt_prb_1}}
\Statex{$\bullet$ \texttt{Encoder}$_\phi$: Pre-trained VAE encoder associated with the inference model $q_\phi(\mathbf{z} | \mathbf{x})$ in \Eqref{vae_inf_dist}}
\Statex{$\bullet$ $\boldsymbol{\mu}_{\phi}(\mathcal{D}_i)$: Empirical posterior mean vector over $\mathcal{D}_i$, defined in \Eqref{mu_Di}}
\Statex{$\bullet$ \texttt{Solver(inputs, labels)}: Continuous piecewise linear regression solver implemented in the \texttt{pwlf} Python library of \citet{jekel2019pwlf}}
\Statex
\Statex{\hspace{-.6cm}\textbf{Algorithm:}}
\State {$\mathbf{p} = \{\,\}$ \Comment{placeholder to store the coordinates of the speech power spectra in the latent subspace}}
\State {$y = \{\,\}$ \Comment{placeholder to store the $f_i$ value associated with $\mathbf{p}$}}
\For{$(\mathbf{x}_n, y_n)$ in $\mathcal{D}_i$}                    
   \State {$\boldsymbol{\mu}_{\phi}(\mathbf{x}_n)= \texttt{Encoder}_\phi(\mathbf{x}_n)$}
   \State {$\mathbf{p}_n = \mathbf{U}_i^\top \big(\boldsymbol{\mu}_{\phi}(\mathbf{x}_n) - \boldsymbol{\mu}_{\phi}(\mathcal{D}_i)\big)$}
   \State {$\mathbf{p}.\texttt{append}(\mathbf{p}_n)$}
   \State {$y.\texttt{append}(y_n)$}
\EndFor
\State{$\mathbf{g}_{\eta_i} = \texttt{Solver}(y, \mathbf{p})$}
\Statex
\Ensure{}
\Statex{$\mathbf{g}_{\eta_i}: \mathbb{R}_+ \mapsto \mathbb{R}^{M_i}$: Estimated piecewise linear regression model to map an $f_i$ value to the coordinates of the speech power spectra in the latent subspace characterized by $\mathbf{U}_i$} 
\end{algorithmic}
\end{algorithm}

\begin{algorithm}[H] 
\caption{Fundamental and formant frequency manipulation in the learned latent subspaces.}
\label{alg:control}
\begin{algorithmic}[1]
\Require{}
\Statex{$\bullet$ $\mathbf{x} \in \mathbb{R}_+^D$: Speech power spectrum}
\Statex{$\bullet$ $y \in \mathbb{R}_+$: Target value for the factor $f_i$ to be modified}
\Statex{$\bullet$ $\mathbf{U}_i \in \mathbb{R}^{L \times M_i}$: Latent subspace matrix associated with $f_i$ obtained by solving \plaineqref{pca_opt_prb_1}}
\Statex{$\bullet$ \texttt{Encoder}$_\phi$: Pre-trained VAE encoder associated with the inference model $q_\phi(\mathbf{z} | \mathbf{x})$ in \Eqref{vae_inf_dist}}
\Statex{$\bullet$ \texttt{Decoder}$_\theta$: Pre-trained VAE decoder associated with the generative model $p_\theta(\mathbf{x} | \mathbf{z})$ in \Eqref{vae_gen_dist}}
\Statex{$\bullet$ $\mathbf{g}_{\eta_i}: \mathbb{R}_+ \mapsto \mathbb{R}^{M_i}$: Regression model obtained with Algorithm~\ref{alg:regression}}
\Statex{$\bullet$ $\texttt{randn}(\boldsymbol{\mu},\boldsymbol{\Sigma})$: Function to sample from a multivariate Gaussian distribution with mean vector $\boldsymbol{\mu}$ and covariance matrix $\boldsymbol{\Sigma}$}
\Statex
\Statex{\hspace{-.6cm}\textbf{Algorithm:}}
\State {$\boldsymbol{\mu}_{\phi}(\mathbf{x}), \mathbf{v}_{\phi}(\mathbf{x}) = \texttt{Encoder}_\phi(\mathbf{x})$}
\State {$\mathbf{z} = \texttt{randn}(\boldsymbol{\mu}_{\phi}(\mathbf{x}_n), \diag\{\mathbf{v}_{\phi}(\mathbf{x})\})$}
\State {$\tilde{\mathbf{z}} = \mathbf{z} - \mathbf{U}_i\mathbf{U}_i^\top \mathbf{z} + \mathbf{U}_i\mathbf{g}_{\eta_i}(y)$}
\State {$\tilde{\mathbf{x}} = \texttt{Decoder}_\theta(\tilde{\mathbf{z}})$}
\Statex
\Ensure{}
\Statex{$\tilde{\mathbf{x}} \in \mathbb{R}_+^D$: Transformed speech power spectrum corresponding to the input target value $y$ for the factor $f_i$}
\end{algorithmic}
\end{algorithm}

\section{Experimental setup details}
\label{app:setup}

\begin{figure*}[t]    
     \centering
     \begin{subfigure}{0.4\textwidth}
         \centering
         \includegraphics[width=0.95\textwidth]{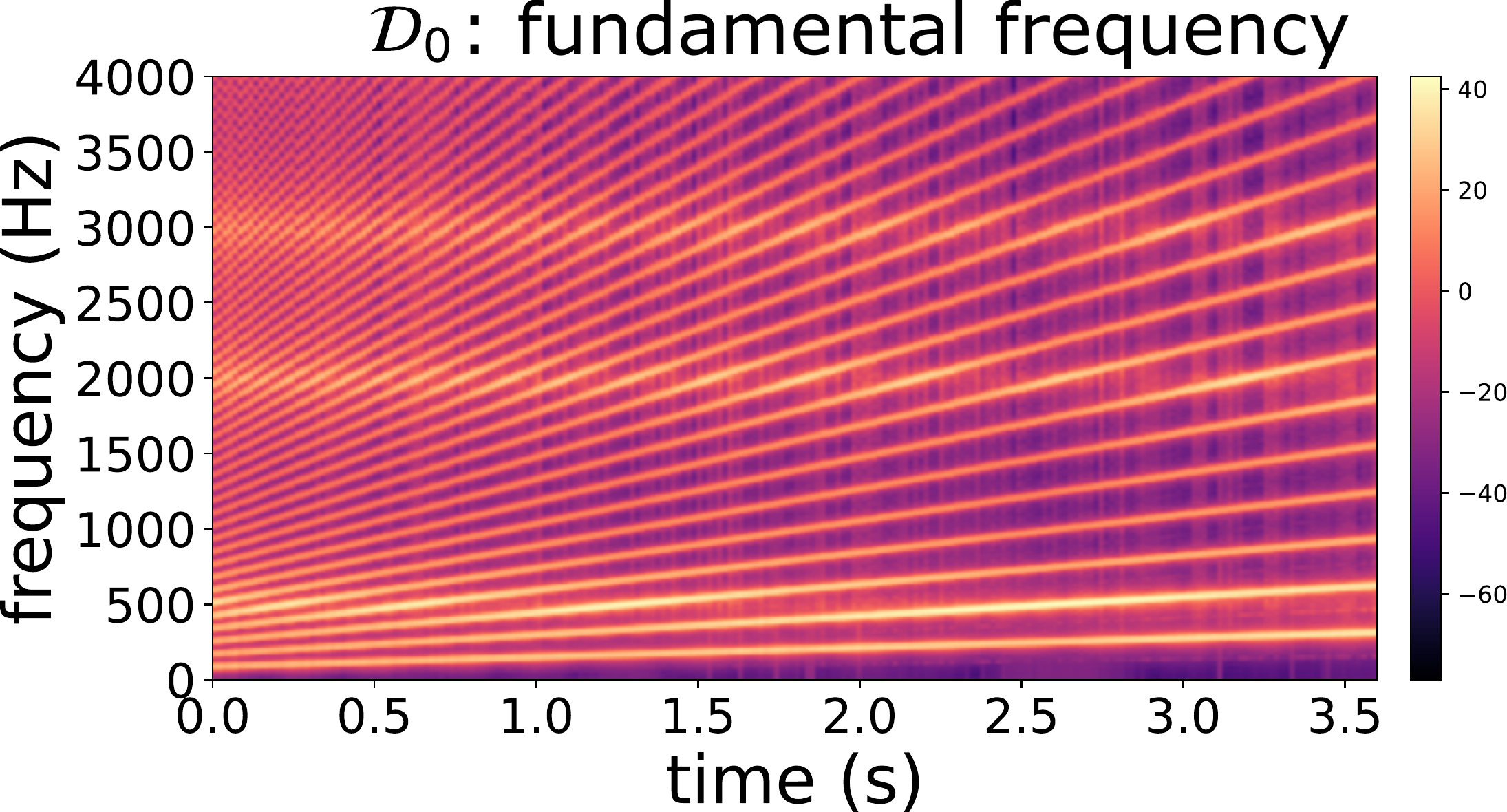}
         \label{fig:f0-trajectory}
     \end{subfigure}
     \begin{subfigure}{0.4\textwidth}
         \centering
         \includegraphics[width=0.95\textwidth]{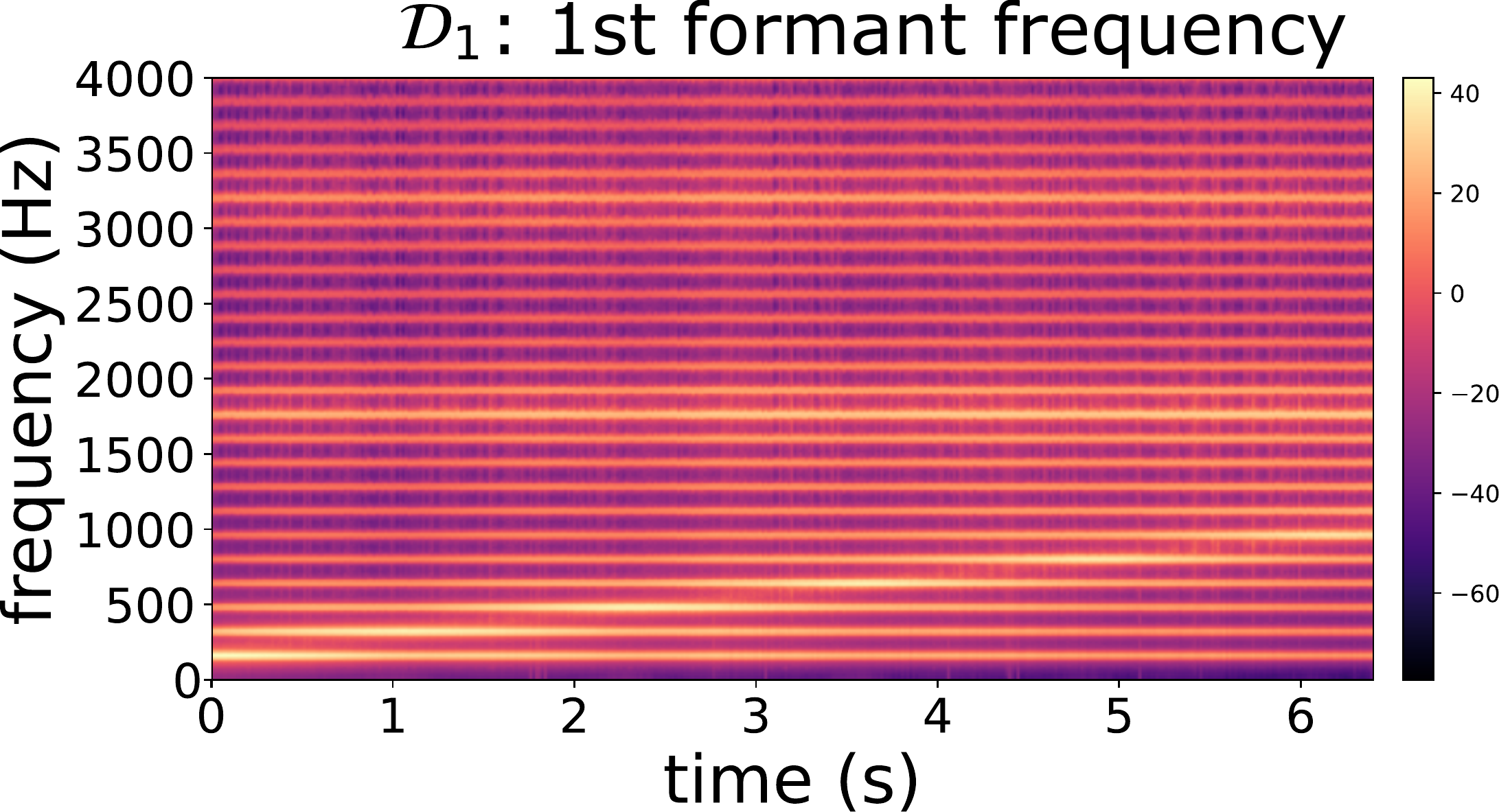}
         \label{fig:f1-trajectory}
     \end{subfigure}
     \begin{subfigure}{0.4\textwidth}
         \centering
         \includegraphics[width=0.95\textwidth]{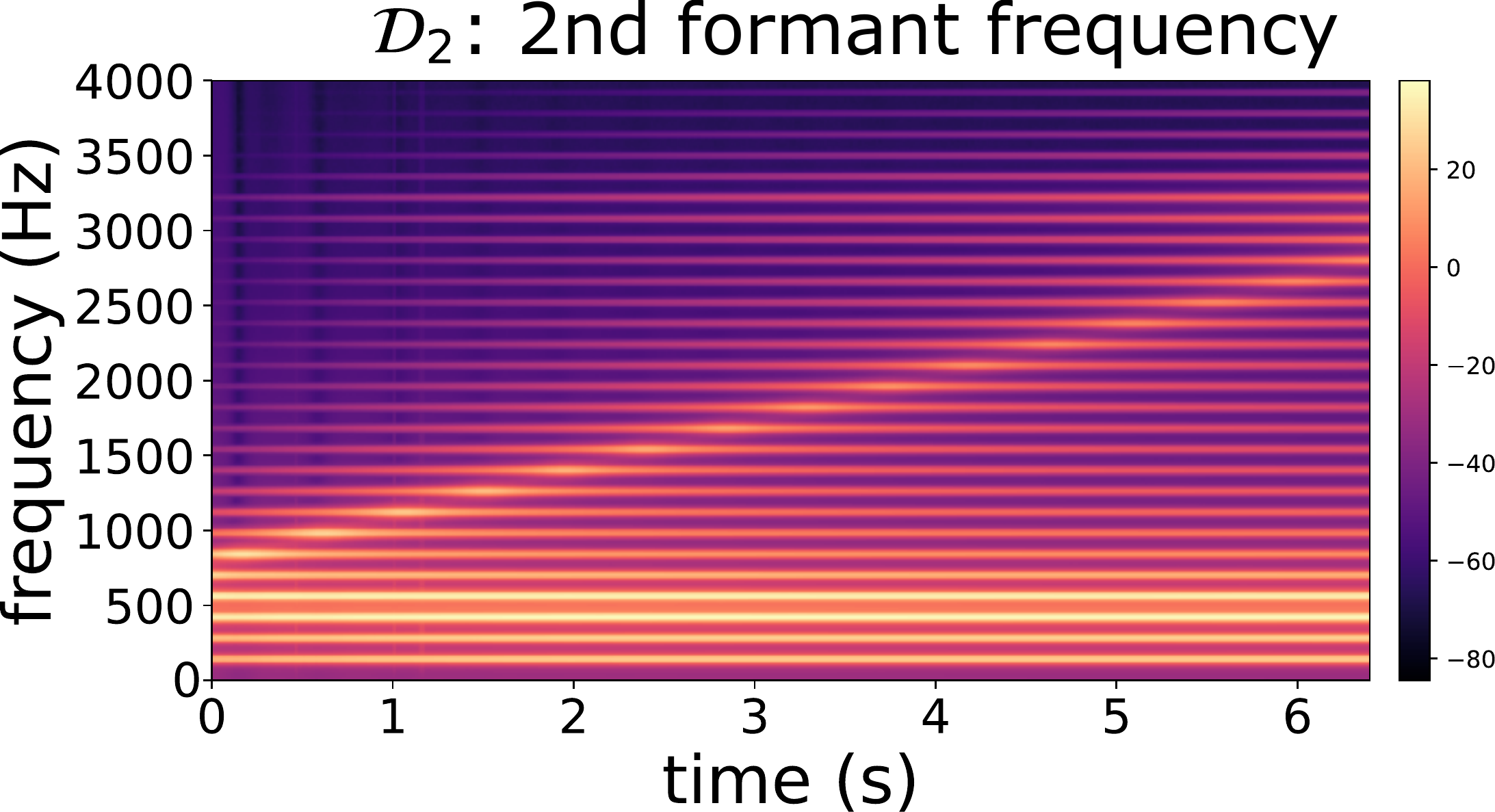}
         \label{fig:f2-trajectory}
     \end{subfigure}
     \begin{subfigure}{0.4\textwidth}
         \centering
         \includegraphics[width=0.95\textwidth]{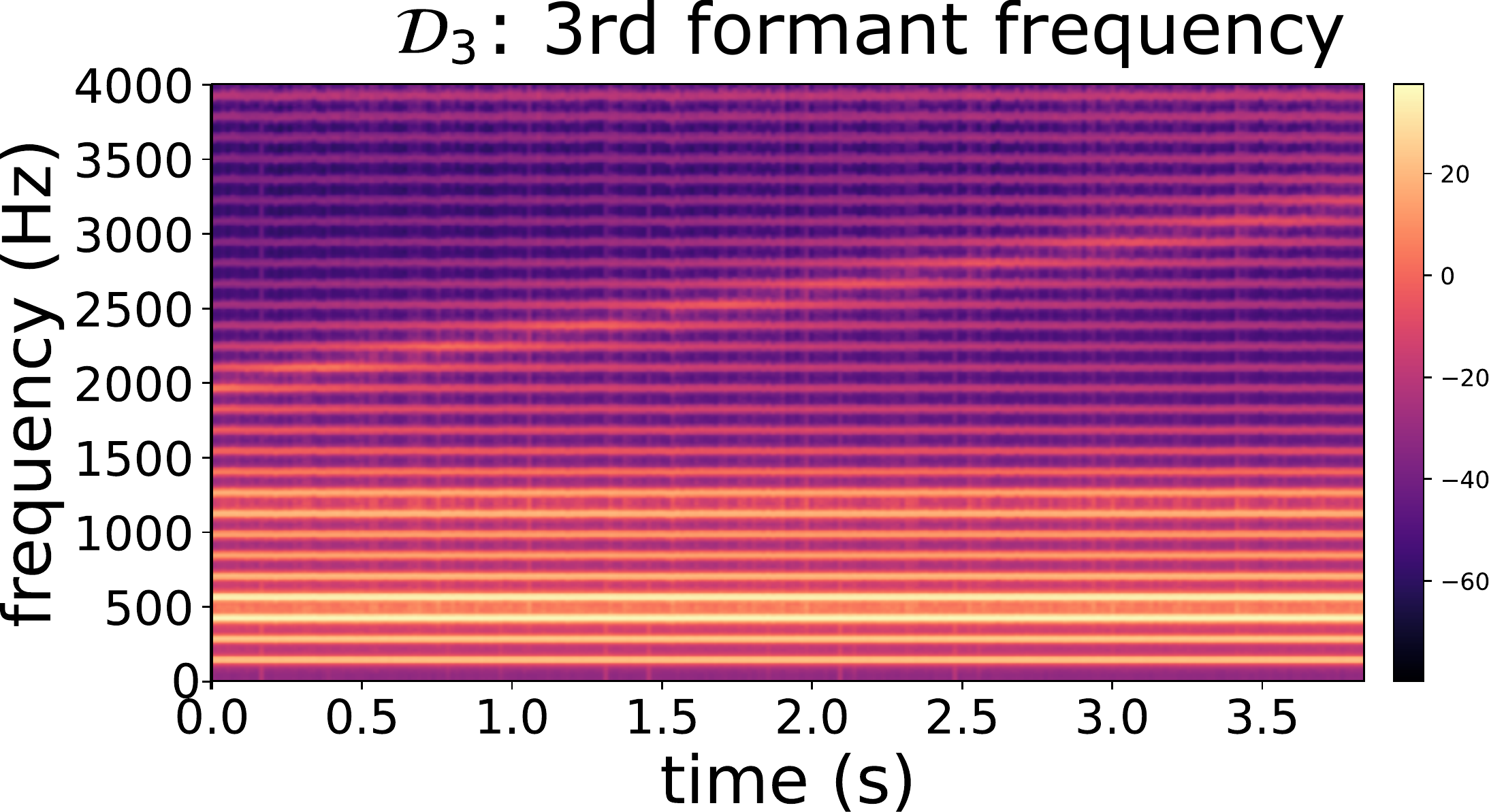}
         \label{fig:f3-trajectory}
     \end{subfigure}
     \caption{Trajectories of speech power spectra generated with Soundgen \citep{Anikin2019}, where only one factor of variation globally varies in each trajectory. From top to bottom and from left to right: the trajectory of the fundamental frequency $f_0$, the trajectory of the formant $f_1$, the trajectory of the formant $f_2$ and the trajectory of the formant $f_3$.}
     \label{fig:trajectories}
\end{figure*}

\paragraph{VAE training} 

To train the IS-VAE model \citep{bando2018statistical,leglaive_MLSP18, girin2019}, we use the Wall Street Journal (WSJ0) dataset \citep{WSJ0}, which contains 25 hours of speech signals sampled at 16 kHz, including 52 female and 49 male speakers. The time-domain speech signals are converted to power spectrograms using the short-time Fourier transform (STFT) with a Hann analysis window of length 64~ms (1,024 samples) and an overlap of 75\%. The VAE input/output dimension is $D=513$ (we only keep the non-redundant part of the power spectrogram corresponding to positive frequencies) and the latent vector dimension is set to $L= 16$. The VAE encoder and decoder networks each have three dense layers. Their dimensions (input dimension, output dimension) are (513, 256), (256, 64) and (64, $2 \times 16$) for the encoder, and (16, 64), (64, 256) and (256, 513) for the decoder\footnote{We have two parameter vectors (mean and variance) for the distribution of $\mathbf{z}$ at the encoder, whereas we have only one single parameter vector (scale) for the distribution of $\mathbf{x}$ at the decoder, see \Eqref{vae_gen_dist} and \Eqref{vae_inf_dist}.}. A hyperbolic tangent ($\tanh$) activation function is used at each layer, except for the output layers of the encoder and decoder where we use the identity function. We train the model using the Adam optimizer \citep{DBLP:journals/corr/KingmaB14} with a learning rate equal to 0.001.

\paragraph{Artificially generated speech data} 

For a given factor of variation, the corresponding latent subspace is learned  using trajectories of speech power spectra generated with Soundgen \citep{Anikin2019}, all other factors being arbitrarily fixed (see Section~\ref{subsec:learning_latent_subspaces}). For $f_0$, the trajectory contains $226$ points (which corresponds to $3.6$ seconds of speech) evenly spaced in the range $[85, 310]$ Hz, $f_1$, $f_2$ and $f_3$ being fixed to $600$ Hz, $2000$ Hz, and $3000$ Hz, respectively. For $f_1$, the trajectory contains $401$ points (which corresponds to $6.4$ seconds of speech) evenly spaced in the range $[200, 1000]$ Hz, $f_0$, $f_2$ and $f_3$ being fixed to $140$ Hz, $1600$ Hz, and $3200$ Hz, respectively. For $f_2$, the trajectory contains $401$ points evenly spaced in the range $[800, 2800]$ Hz, $f_0$, $f_1$ and $f_3$ being fixed to $140$ Hz, $500$ Hz, and $3200$ Hz, respectively. For $f_3$, the trajectory contains $241$ points (which corresponds to $3.9$ seconds of speech) evenly spaced in the range $[2000, 3200]$ Hz, $f_0$, $f_1$ and $f_2$ are fixed to $140$ Hz, $500$ Hz, and $1200$ Hz, respectively. These four trajectories are illustrated in Figure~\ref{fig:trajectories}.
The amplitude of the formants is fixed at $30$dB, and their bandwidth is automatically calculated from the formant frequencies using a formula derived from phonetics studies. Quoting the documentation of Soundgen \citep{Anikin2019}, ``above $500$~Hz [the bandwidth] follows the original formula known as ``TMF-1963'' \citep{tappert1963spectrum}, and below $500$ Hz it applies a correction to allow for energy losses at low frequencies \citep{khodai2002comparative}. Below $250$~Hz the bandwidth starts to decrease again, in a purely empirical attempt to achieve reasonable values even for formant frequencies below ordinary human range. See the internal function soundgen:::getBandwidth().''
The regression models used to control the speech factors of variation in the latent space (see Section~\ref{subsec:control_factor}) are learned on the same trajectories, but using the values of the Soundgen input parameters.


\section{Visualization of the learned latent subspaces}
\label{app:viz}

\begin{table}[h]
\centering
\caption{Cumulative variance (in $\%$) retained by the projection $\mathbf{U}_i\mathbf{U}_i^\top$, $\mathbf{U}_i \in \mathbb{R}^{L \times M_i}$, as a function of the number of components $M_i$. We keep as much components as needed to retain at least 80 $\%$ of the data variance, as indicated by the underlined numbers.}
\label{tab:variance}
\begin{tabular}{l|llll}
     & $f_0$ & $f_1$ & $f_2$ & $f_3$ \\ \hline
    $M_i = 1$ component &   33 &  \underline{81} &  39 &  48 \\
    $M_i = 2$ components &  59 &    87  &  70 &  73 \\
    $M_i = 3$ components &  78 &        90         &  \underline{88} &  \underline{83} \\
    $M_i = 4$ components &  \underline{90} & 92 &  93 & 87 \\
\end{tabular}
\end{table}

\begin{figure}[!ht]
     \centering
     \begin{subfigure}[b]{0.33\linewidth}
         \centering
         \includegraphics[width=\linewidth]{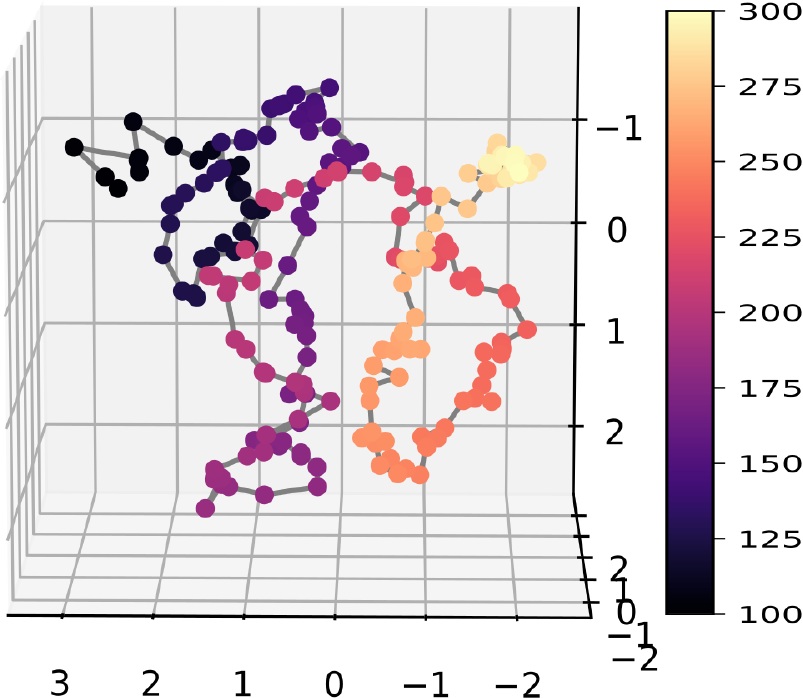}
         \caption{$f_0$ trajectory corresponding to $\mathcal{D}_0$.}
        \label{fig_viz_f0}
     \end{subfigure} \hspace{1cm}
     \begin{subfigure}[b]{0.33\linewidth}
         \centering
         \includegraphics[width=\linewidth]{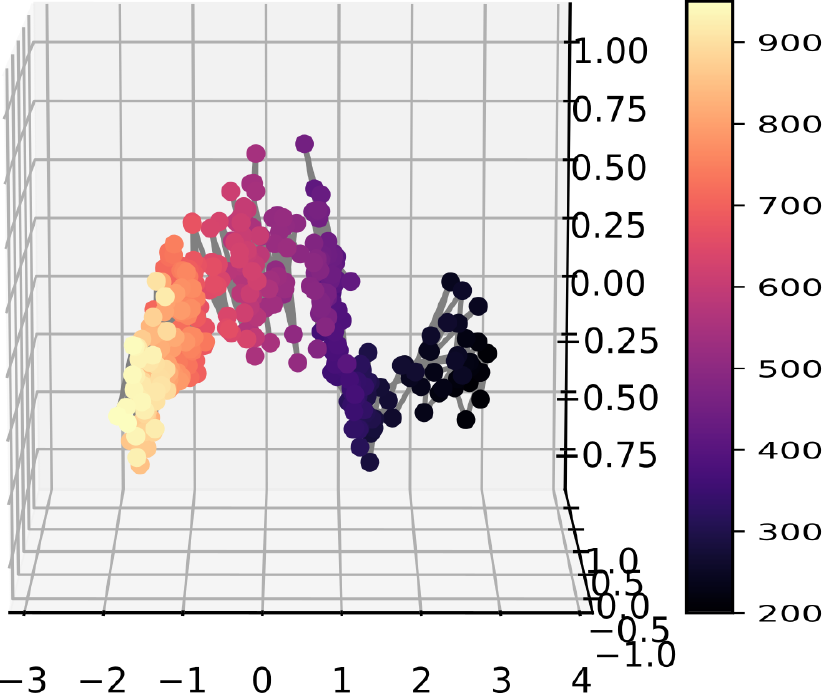}
         \caption{$f_1$ trajectory corresponding to $\mathcal{D}_1$.}
         \label{fig_viz_f1}
     \end{subfigure} \hspace{1cm}
     \begin{subfigure}[b]{0.33\linewidth}
         \centering
         \includegraphics[width=\linewidth]{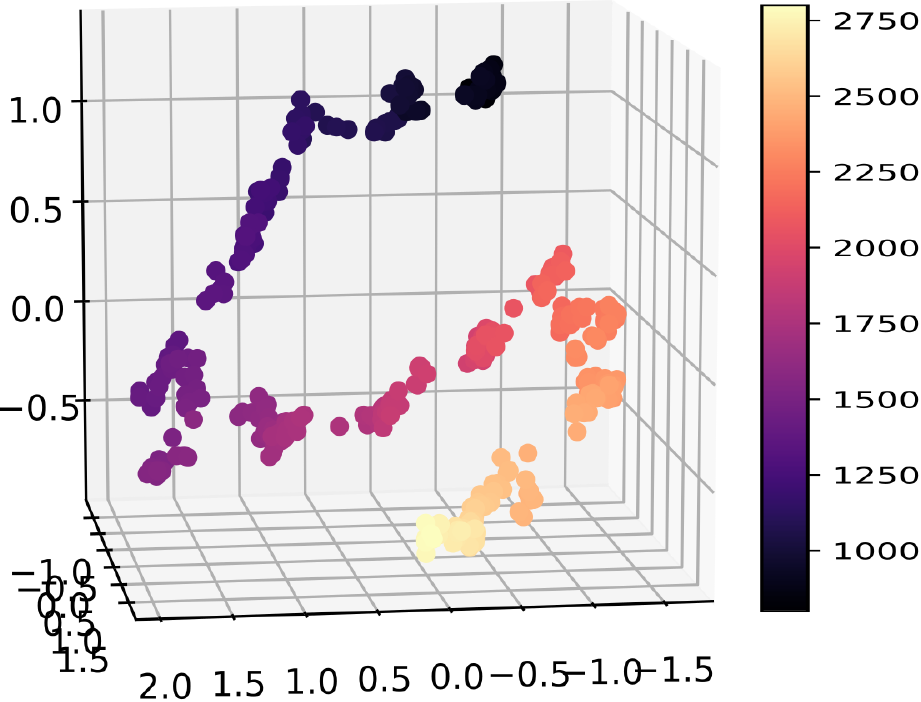}
         \caption{$f_2$ trajectory corresponding to $\mathcal{D}_2$.}
         \label{fig_viz_f2}
     \end{subfigure} \hspace{1cm}
     \begin{subfigure}[b]{.33\linewidth}
         \centering
         \includegraphics[width=\linewidth]{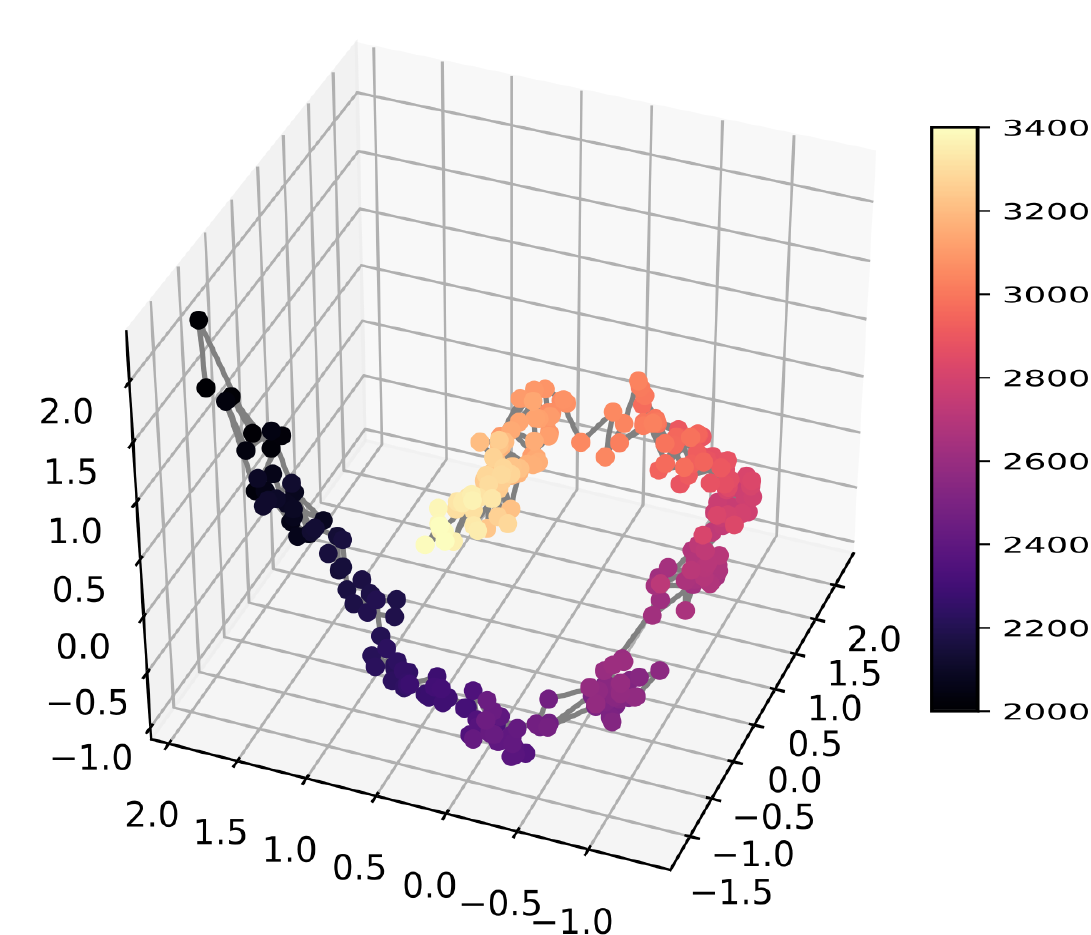}
         \caption{$f_3$ trajectory corresponding to $\mathcal{D}_3$.}
         \label{fig_viz_f3}
     \end{subfigure} \hspace{1cm}
     \begin{subfigure}[b]{.3\linewidth}
         \centering
         \includegraphics[width=\linewidth]{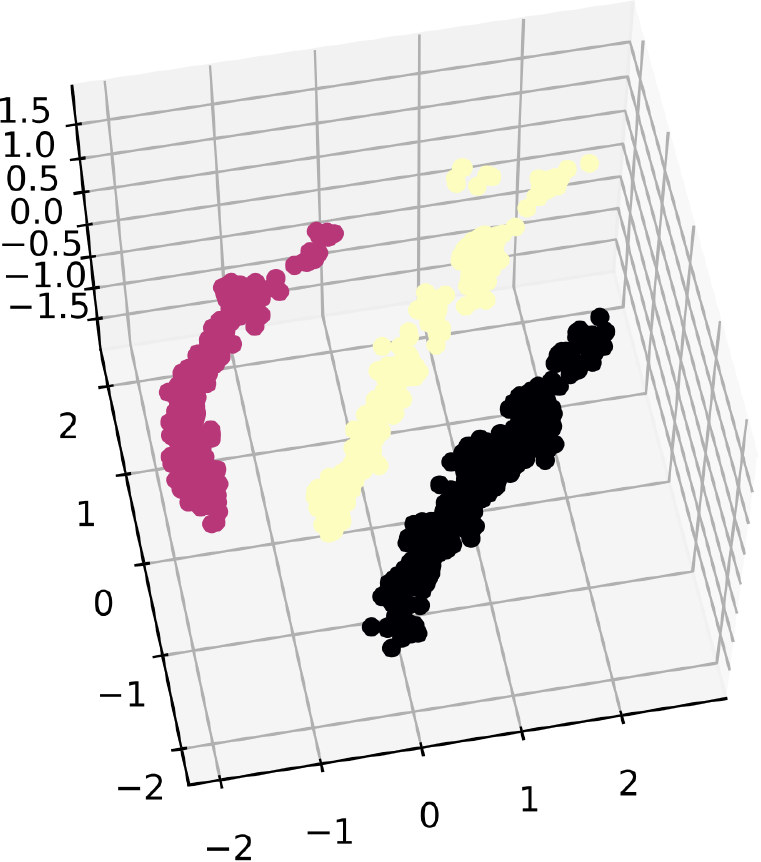}
         \caption{Three trajectories for $f_1$, associated with different values of $f_2$.}
         \label{fig_viz_f1_multi}
     \end{subfigure} \hspace{1cm}
     \begin{subfigure}[b]{.3\linewidth}
         \centering
         \includegraphics[width=\linewidth]{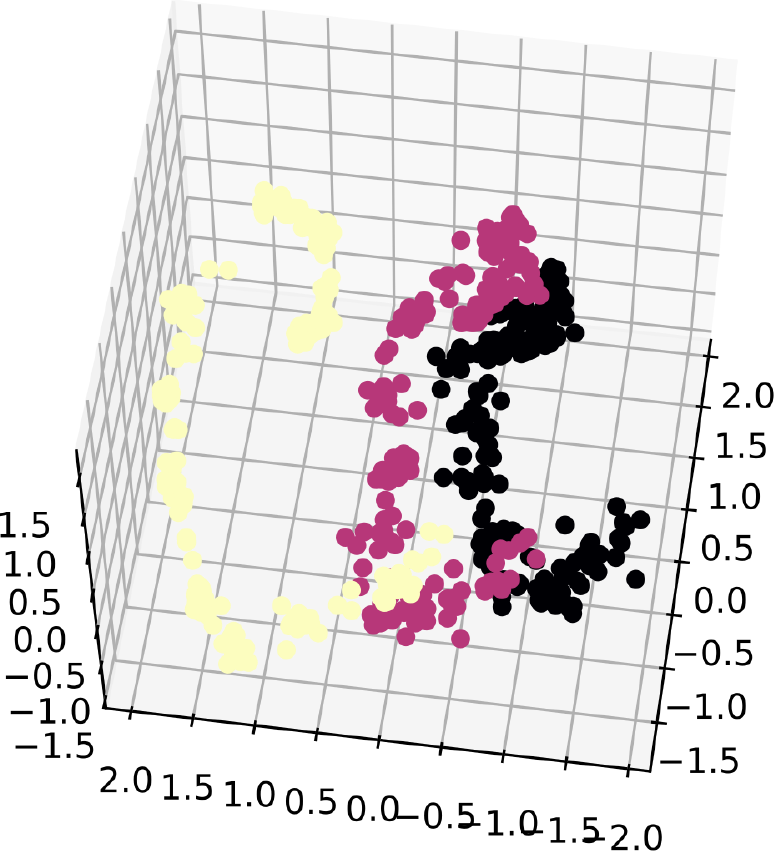}
         \caption{Three trajectories for $f_2$, associated with different values of $f_1$.}
         \label{fig_viz_f2_multi}
     \end{subfigure}
    \caption{Visualization of trajectories in the learned latent subspaces.}
    \label{fig:viz}
\end{figure}

For $i = 0, 1, 2$ and $3$, Figures~\ref{fig_viz_f0}, \ref{fig_viz_f1}, \ref{fig_viz_f2} and \ref{fig_viz_f3} are respectively obtained by projecting the latent mean vectors $\boldsymbol{\mu}_\phi(\mathbf{x}) \in \mathbb{R}^L$, for all data vectors $\mathbf{x} \in \mathcal{D}_i$, within the latent subspace characterized by $\mathbf{U}_i \in \mathbb{R}^{L \times M_i}$ (i.e., we perform dimensionality reduction). In the reported experiments, the latent subspace dimension $M_i$ for each factor of variation was chosen such that $80\%$ of the data variance was retained in the latent space. As indicated in Table~\ref{tab:variance}, this resulted in $M_0 = 4$, $M_1 = 1$ and $M_2 = M_3 = 3$. In this section, for visualization purposes, we set $M_i = 3$ for all $i \in \{0,1,2,3\}$. However, we can see that the $f_1$ trajectory (Figure~\ref{fig_viz_f1}) is mainly concentrated along a single axis, as indicated by the amount of variance retained by this axis 81\% (see Table~\ref{tab:variance}). Regarding $f_0$ (Figure~\ref{fig_viz_f0}), setting $M_0 = 3$ retained 78$\%$ of the variance of $\mathcal{D}_0$ in the latent space. From these figures, we see that two data vectors $\mathbf{x}$ and $\mathbf{x}'$ corresponding to two close values of a given factor have projections of $\boldsymbol{\mu}_\phi(\mathbf{x})$ and $\boldsymbol{\mu}_\phi(\mathbf{x}')$ that are also close in the learned latent subspaces. This can be seen from the color bars which indicate the values of the factors of variation. The learned representation thus preserves the notion of proximity in terms of $f_0$ and formant frequencies.

In Figure~\ref{fig_viz_f1_multi}, we project three different datasets $\mathcal{D}_1$, defined for three different values of $f_2$. Similarly, in Figure~\ref{fig_viz_f2_multi} we show the trajectories associated with the projection of three datasets $\mathcal{D}_2$, defined for three different values of $f_1$. We notice that as expected, the trajectories are very similar and mostly differ by a translation.


\section{Correlation matrices obtained from MFCCs and short-term magnitude spectra}
\label{app:correlation_mfcc_spec}
We conducted experiments with the proposed method (i.e., learning the subspace for each factor of variation and then learning a regression model to move in the learned subspace) on the following representations: MFCC and short-term magnitude spectrum (i.e., columns of the STFT magnitude spectrogram). We used the same artificial dataset as for the VAE latent space representation. Figures~\ref{fig:mfcc_correlation_} and~\ref{fig:spectrogram_correlation_} below show the correlation matrix of the latent subspace basis vectors learned for the MFCC and short-term magnitude spectrum, respectively. Similarly to what we did with the VAE, the dimension of the subspaces is determined by applying a threshold on the data variance ($\geq $80\%). For the dimension of the $f_0$ subspace with the MFCCs, we actually need 15 components to keep 80\% of the data variance, so here we take only the principal one to facilitate the reading.

\clearpage

\begin{figure}[!ht]
     \centering
     \begin{subfigure}[b]{0.45\textwidth}
         \centering
         \includegraphics[width=\textwidth]{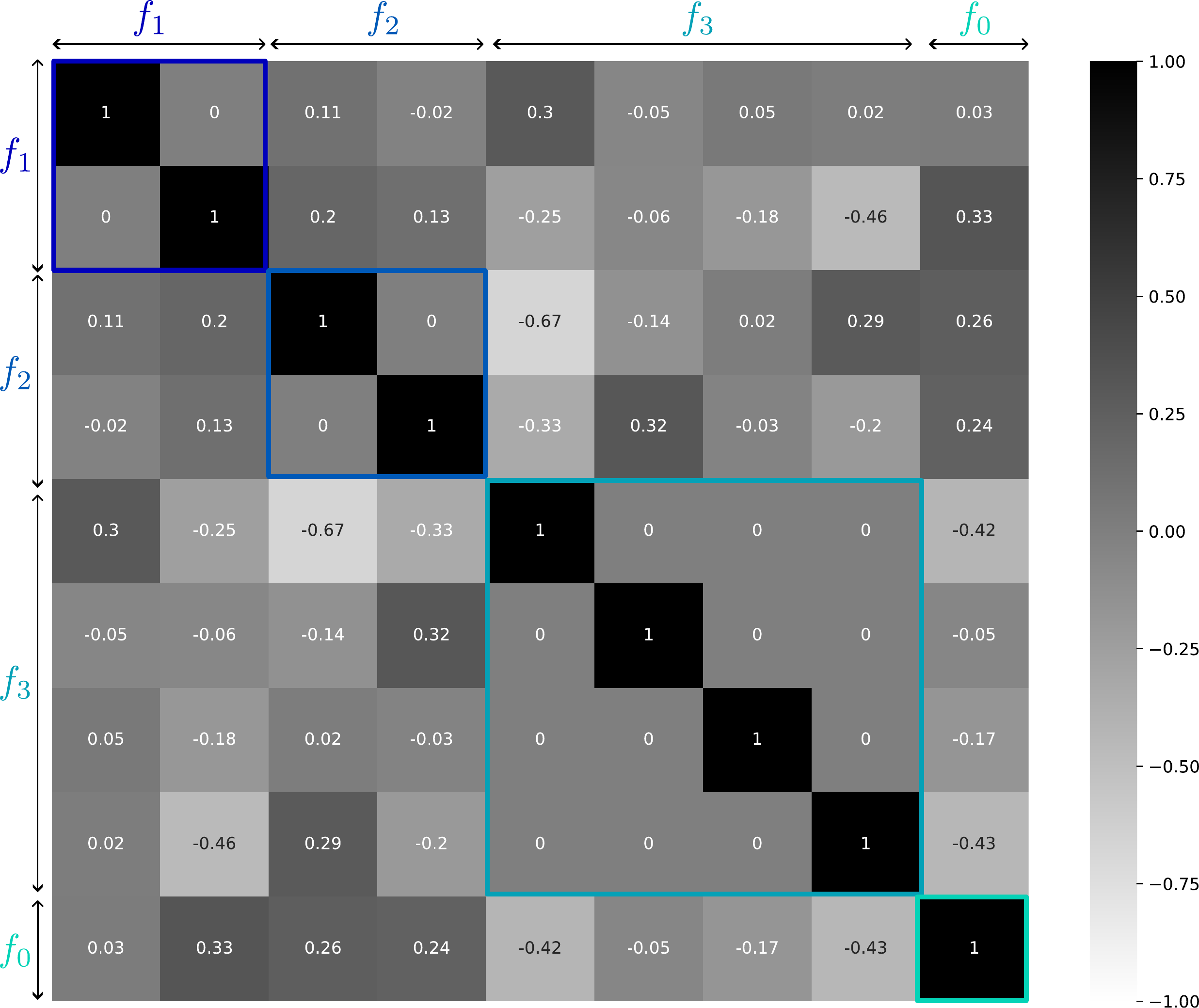}
         \caption{MFCC}
         \label{fig:mfcc_correlation_}
     \end{subfigure}
     \begin{subfigure}[b]{0.45\textwidth}
         \centering
         \includegraphics[width=\textwidth]{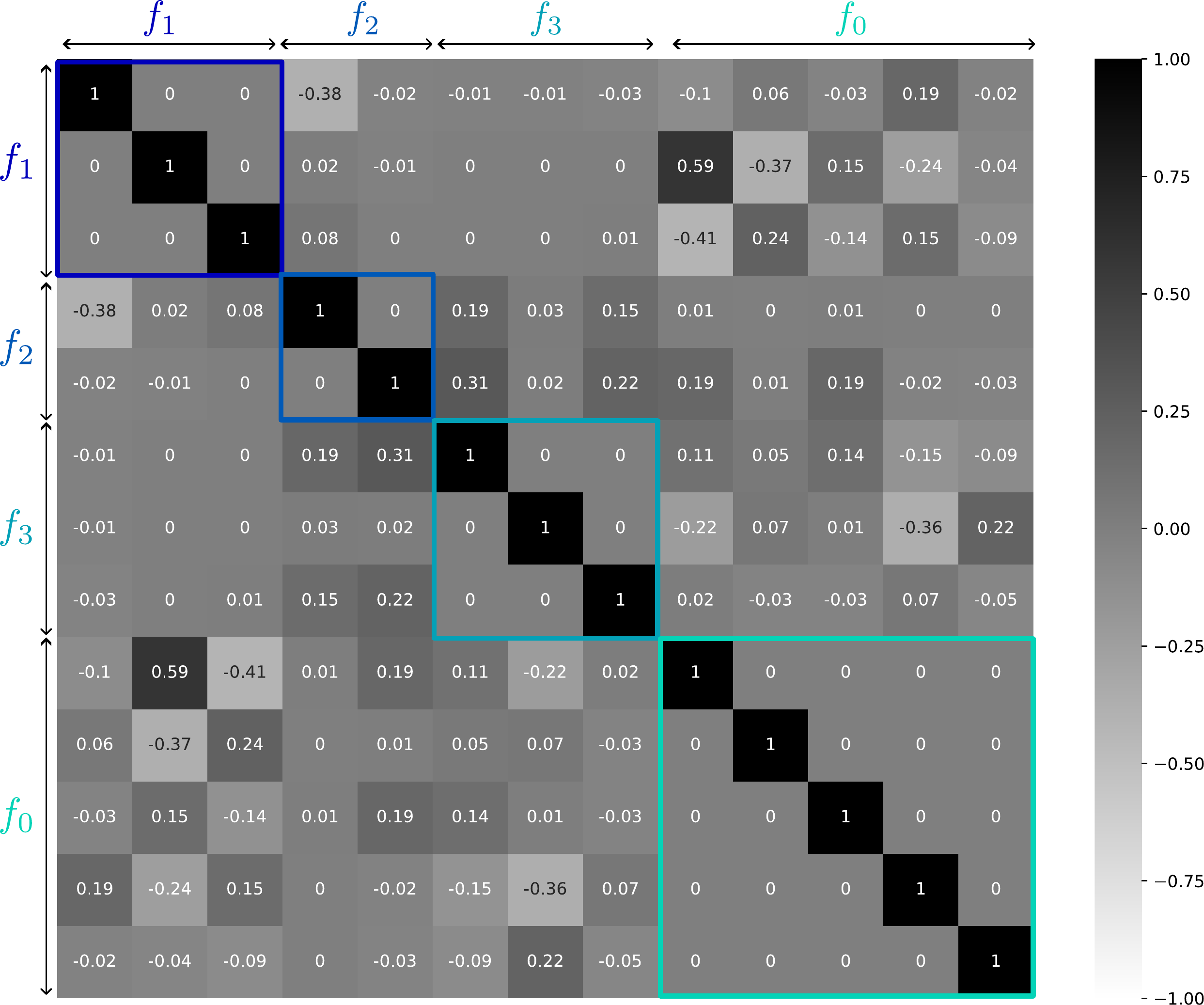}
         \caption{short-term magnitude spectrum}
         \label{fig:spectrogram_correlation_}
     \end{subfigure}
     
     \caption{Correlation matrix of the latent subspace basis vectors learned for MFCC and short-term magnitude spectrum.}
     \label{fig:other-representations_}
\end{figure}

\bibliography{mybibfile}
\balance

\end{document}